\documentclass[pdflatex,sn-mathphys-num]{sn-jnl}% Math and Physical Sciences Numbered Reference Style
\usepackage{graphicx}%
\usepackage{multirow}%
\usepackage{amsmath,amssymb,amsfonts}%
\usepackage{amsthm}%
\usepackage{mathrsfs}%
\usepackage[title]{appendix}%
\usepackage{xcolor}%
\usepackage{textcomp}%
\usepackage{manyfoot}%
\usepackage{booktabs}%
\usepackage{algorithm}%
\usepackage{algorithmicx}%
\usepackage{algpseudocode}%
\usepackage{listings}%
\usepackage{subcaption}
\usepackage{url}
\theoremstyle{thmstyleone}%
\theoremstyle{thmstyletwo}%

\theoremstyle{thmstylethree}%

{\begin{tableorg}[#1]\begin{center}}
{\end{center}\end{tableorg}}

\begin{document}

% \title[Article Title]{Explainable and Generalisable LLM-based Alzheimer's Detection with Spontaneous Speech}
\title[Article Title]{Explainable and Generalisable LLM-based Cognitive Decline Detection with Spontaneous Speech}

\author[1,2]{\fnm{Ziyun} \sur{Cui}}\email{cui-zy24@mails.tsinghua.edu.cn}
\author[2]{\fnm{Wen} \sur{Wu}}\email{wuwen@pjlab.org.cn}
\author[3,4]{\fnm{Chuan} \sur{Shi}}\email{shichuan@bjmu.edu.cn}
\author[3]{\fnm{Shuguang} \sur{Yang}}\email{2311210597@bjmu.edu.cn}
\author[5]{\fnm{Xueying} \sur{Gui}}\email{guixueying@oppo.com}
\author[5]{\fnm{Yan} \sur{Zheng}}\email{zhengyan1@oppo.com}
\author[6]{\fnm{Qiong} \sur{Yang}}\email{yangqiongputh@126.com}
\author[6]{\fnm{Haiyan} \sur{Zhao}}\email{hailan200159@sina.cn}
\author[1]{\fnm{Wei-Qiang} \sur{Zhang}}\email{wqzhang@tsinghua.edu.cn}
\author[1]{\fnm{Ji} \sur{Wu}}\email{wuji\_ee@tsinghua.edu.cn}
\author[5]{\fnm{Yelei} \sur{Li}}\email{liyelei1@oppo.com}
\author*[4,6]{\fnm{Nan} \sur{Li}}\email{linan917@163.com}
\author*[1,2]{\fnm{Chao} \sur{Zhang}}\email{cz277@tsinghua.edu.cn}

\affil*[1]{\orgname{Tsinghua University}, \orgaddress{\city{Beijing}, \country{China}}}

\affil[2]{\orgname{Shanghai Artificial Intelligence Laboratory}, \orgaddress{\city{Shanghai}, \country{China}}}

\affil[3]{\orgname{Peking University Sixth Hospital}, \orgaddress{\city{Beijing}, \country{China}}}

\affil[4]{\orgname{Peking University}, \orgaddress{\city{Beijing}, \country{China}}}

\affil[5]{\orgname{OPPO Research Institute}, \orgaddress{\city{Shenzhen}, \country{China}}}

\affil[6]{\orgname{Peking University Third Hospital}, \orgaddress{\city{Beijing}, \country{China}}}


% \abstract{The abstract serves both as a general introduction to the topic and as a brief, non-technical summary of the main results and their implications. Authors are advised to check the author instructions for the journal they are submitting to for word limits and if structural elements like subheadings, citations, or equations are permitted.}
\abstract{
  Alzheimer’s disease (AD) and mild cognitive impairment (MCI), which may precede AD, manifest early through subtle linguistic and acoustic alterations. Traditional diagnostics, however, are often resource-intensive and lack scalability for mass screening. To address these challenges, we introduce a novel bilingual speech large language model framework for automated, explainable cognitive screening. Unlike conventional pipelines that rely on error-prone automatic speech recognition, our system directly processes raw speech to learn joint acoustic-semantic representations, preserving critical prosodic cues often lost in transcription. Utilising our newly collected PUTH-AD dataset alongside multiple open-source corpora, we implemented a multi-task learning objective that simultaneously performs cognitive status classification and generates clinician-understandable natural language explanations. 
  Our system achieved the highest average accuracy and AUROC across six dataset/task conditions, comparing three representative baselines. 
  The system demonstrated cross-task transfer to held-out PUTH-AD task subsets, maintaining classification accuracy on an entirely unseen cognitive task without task-specific fine-tuning.
  Furthermore, clinician evaluation confirms that the generated explanations are both clinically relevant and largely consistent with the underlying speech evidence, supporting their potential utility in clinical interpretation. 
  This study provides a scalable, objective, and explainable framework for speech-based cognitive screening, combining cognitive status classification with natural language explanations that clinicians can assess and verify, bridging the gap between advanced AI and clinical utility.
  }




\maketitle

\renewcommand{\figurename}{Figure}

\section*{Introduction}
Alzheimer’s disease (AD) is a neurodegenerative disorder characterised by a long-term and progressive decline in cognitive functioning~\cite{deture2019neuropathological}. 
Early dentification of AD and mild cognitive impairment (MCI), which may precede AD, is critical for timely intervention, care planning, and potential disease-modifying treatment~\cite{nestor2004advances, crous2017alzheimer}. 
Conventional diagnostic pathways rely on clinical interviews and neuropsychological assessments, which are time-consuming, resource-intensive, and require trained specialists~\cite{knopman2000patterns, porsteinsson2021diagnosis}. These limitations hinder large-scale screening and longitudinal monitoring, motivating the development of scalable and cost-effective alternatives.
Speech has emerged as a promising low-cost digital biomarker for cognitive decline. Unlike neuroimaging or biomarker assays, speech can be collected non-invasively using ubiquitous devices such as smartphones or telehealth platforms, enabling frequent, remote, and population-scale assessment with minimal burden on patients and healthcare systems. As a naturally occurring behaviour tightly coupled with cognitive and linguistic processing, speech reflects subtle impairments in memory, executive function, and language planning. Notably, speech dysfunction is an early symptom of AD, manifesting as increased pauses and hesitations, word-finding difficulties, repetitions, reduced lexical diversity, overuse of vague expressions, and inappropriate pronoun usage~\cite{cummings2019describing}.
A growing body of research demonstrates that AD-related impairment can be detected from acoustic and linguistic characteristics of speech. These include hand-crafted acoustic features such as short-time energy and spectral centroid~\cite{lopez2012new}, fluency-related measures~\cite{konig2015automatic}, spectrogram-based representations~\cite{bertini2022automatic,chen2021automatic}, deep-learning-derived acoustic features~\cite{koo2020exploiting,zhu2021wavbert,chen2023cross}, as well as linguistic representations including GloVe embeddings~\cite{rohanian2021alzheimer} and features extracted using self-supervised learning (SSL) models~\cite{agbavor2022predicting}. Together, these advances highlight spontaneous speech as an accessible, scalable, and clinically meaningful modality for automated AD screening and monitoring.

Large language models (LLMs) are Transformer-based generative models trained on massive corpora, which enable them with strong generalisation, in-context learning and instruction-following abilities. Recent models extend beyond text to multimodal inputs (\textit{e.g.}, SpeechLLM, which accepts speech as input), making them well-suited for speech understanding tasks~\cite{tang2023salmonn, chu2024qwen2, ding2025kimi}. With the success of LLMs, interest in using LLMs for speech-based AD detection has grown quickly. Existing work centres on several LLM-driven paradigms: (i) fine-tuning an LLM and feeding its embedding, serving as speech feature, to downstream classifiers for end-to-end discriminative modelling~\cite{casu2025integrating, park2025reasoning}; (ii) prompting an LLM to generate macro-descriptors as handcrafted features, and then classifying with these generated features~\cite{li2024devising, heitz2025linguistic, mo2025dect, botelho2024macro}; (iii) employing prompting strategies, such as few-shot and chain-of-thought (CoT), to elicit a direct judgment from LLM~\cite{guo2024interpretable, bt2024performance}. As for interpretability, some studies apply ANOVA and SHAP to quantify the importance of different generated features and to probe their correspondence to cognitive-impairment markers~\cite{li2024devising, heitz2025linguistic}, while others design prompts that lead the model to produce readable free-text rationales, thereby providing a clinician-facing chain of evidence beyond the final decision~\cite{guo2024interpretable}. 

However, existing approaches face several key limitations that hinder their clinical applicability. Firstly, there exists a trade-off between performance and interpretability in current speech-based AD systems. High-performing models typically operate as ``black boxes'' offering limited interpretability, while approaches that generate readable free-text rationales achieve lower classification accuracy compared to other methods on the same datasets. Secondly, most existing systems rely heavily on automatic speech recognition (ASR) to transcribe audio into text before extracting linguistic features or prompting LLMs, which introduces error propagation to the pipeline. Previous studies~\cite{park2025reasoning, heitz2025linguistic} have shown that classification accuracy with ASR transcripts are markedly lower to manual reference transcriptions, which are unavailable in real-world deployment scenarios. Moreover, most recent-year speech-based AD studies~\cite{li2024devising, guo2024interpretable, heitz2025linguistic, casu2025integrating, mo2025dect, park2025reasoning} focus on ADReSS~\cite{luz2020alzheimer} and ADReSSo~\cite{luz2021detecting} datasets, two English-only corpora restricted to the single Cookie Theft task for binary classification of healthy controls (HC) versus AD, limiting clinical granularity as well as system generalisation across languages and tasks.

\begin{figure}[tbh]
    \centering
    \includegraphics[width=0.95\linewidth]{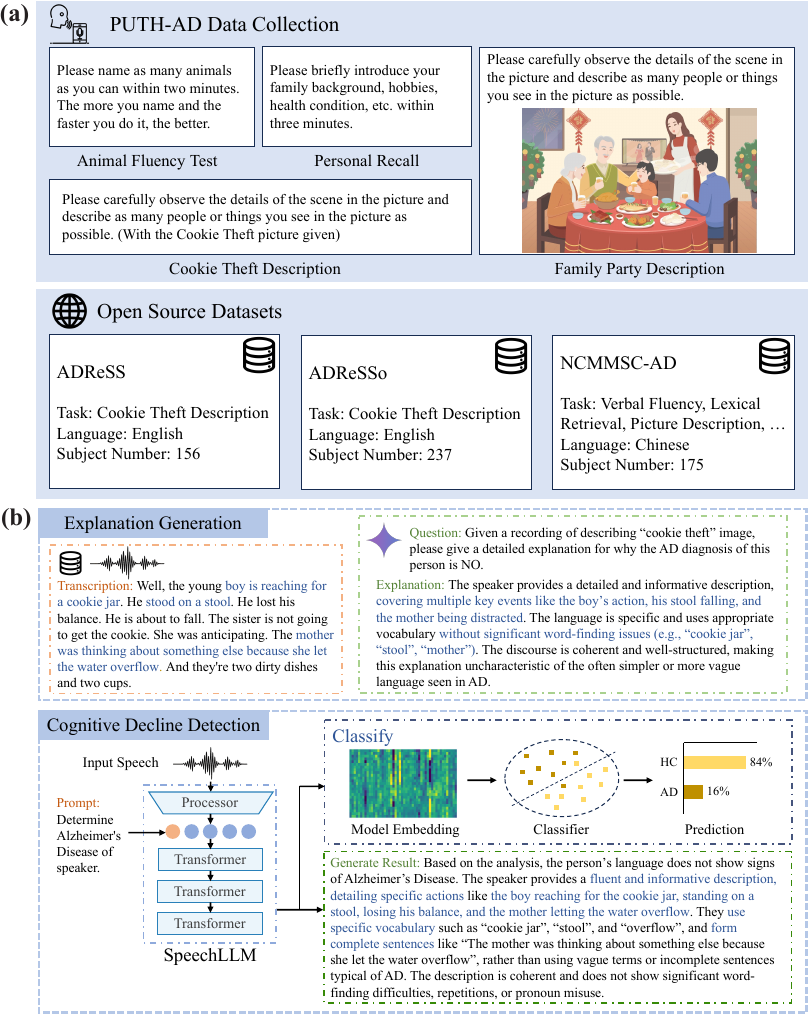}
    \caption{Overview of the proposed bilingual speech-based cognitive screening framework. (a) Dataset composition includes a newly collected PUTH-AD dataset, completing four distinct cognitive tasks, alongside established open-source datasets covering English and Chinese languages with varying task types. (b) The system pipeline consists of two components: 
    Evidence text generation uses Gemini to generate explanations from speech recordings and cognitive status labels, serving as training targets; the proposed SpeechLLM-based system processes raw speech with task prompts to simultaneously output cognitive status classification and natural language explanations.
    }
    \label{fig:overall}
\end{figure}

To address these limitations, we collected the PUTH-AD dataset at Peking University Third Hospital and Peking University Sixth Hospital, in Beijing, China, comprising 166 participants who each completed four distinct tasks: Animal Fluency Test (AFT, listing as many animal names as possible), Cookie Theft Description (CTD, cookie theft picture adapted with Chinese text), Family Party Description (FPD, depicting a Chinese New Year dinner scene), and Personal Recall (PR, three-minute introduction covering family situation, hobbies, and health status). 
Each participant underwent cognitive screening using the MoCA and HKBC scales, yielding score-defined categories of normal cognition, mild cognitive impairment, and dementia-level impairment.
We developed a novel bilingual SpeechLLM-based system that directly processes raw speech without requiring ASR transcription, learning joint acoustic-semantic representations that preserve prosodic cues such as pause patterns, speech rate, and intonation alongside linguistic content. The system employs a multi-task learning framework that simultaneously performs cognitive status classification and explanation generation: a classification head outputs probability distributions over cognitive status categories, while a generation head produces natural language summaries articulating the salient speech characteristics that inform prediction, such as hesitations, word-finding difficulties, or reduced semantic coherence. By jointly optimising both objectives, the model learns representations that are both discriminative for classification and interpretable through clinician-understandable rationales, effectively balancing accuracy and transparency.

Our contributions advance the field of automated cognitive screening in several key aspects. First, we demonstrate that SpeechLLM-based approaches outperform traditional SSL-based pipelines and general-purpose language model baselines across diverse datasets and languages, with multi-task learning enhancing rather than compromising classification accuracy. Second, through cross-domain joint training experiments spanning ADReSS, ADReSSo, NCMMSC-AD, and PUTH-AD datasets, we establish robust generalisation capabilities to speech tasks, including completely unseen speech tasks, addressing the limited scope of prior work focused on single-language, single-task scenarios. Third, our system provides interpretable support through automatically generated explanations that describe clinically relevant speech features, moving beyond post-hoc interpretability techniques to integrate explain ability directly into the model's decision-making process. Fourth, through clinician evaluation, we demonstrate that the generated explanations are both clinically relevant and largely consistent with the underlying speech evidence, providing direct evidence for their potential utility in supporting clinical interpretation. 
These advances collectively offer a scalable, objective, and transparent solution for early dementia screening.

The remainder of this paper is organised as follows. The Results section reports dataset demographics, overall performance and baseline comparisons, ablation results for multi-task learning, generalisation under joint training across multiple datasets, and clinical evaluation of classification and explanation outputs. The Discussion section discusses the main findings and limitations. The Methods section describes the datasets and speech tasks, model architecture, baseline models, statistical analysis, and implementation details.

\section*{Results}
\label{sec: results}

\subsection*{Dataset Statistics}

Our study encompasses three distinct speech corpora spanning English and Chinese languages, collected from both research challenges and clinical settings. The PUTH-AD dataset, collected at Peking University Third Hospital and Peking University Sixth Hospital, consists of 166 participants who all completed four speech assessment tasks: animal fluency testing (AFT), Family Party picture description (FPD), Cookie Theft picture description (CTD), and personal recall (PR). 
The data from ADReSS~\cite{luz2020alzheimer} and ADReSSo~\cite{luz2021detecting} challenges are merged into the ADReSS corpus, including 330 English-speaking participants performing Cookie Theft picture description tasks.
The NCMMSC-AD dataset, from a challenge at the NCMMSC 2021 conference, includes 175 Chinese-speaking participants completing multiple cognitive tasks, including verbal fluency, lexical retrieval, picture description, self-introduction, \textit{etc.}

\begin{figure}
    \centering
    \includegraphics[width=0.95\linewidth]{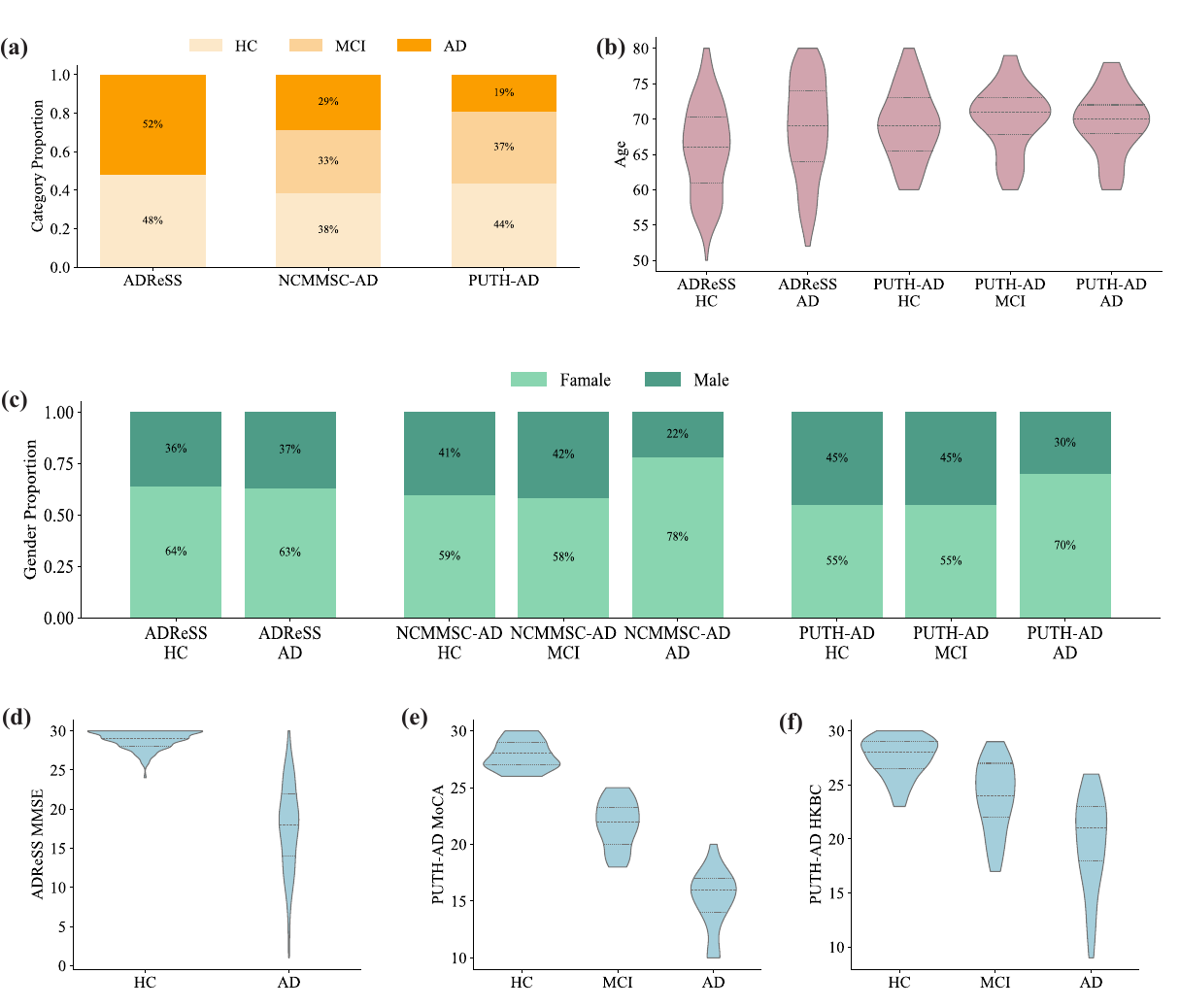}
    \caption{Demographic and cognitive status characteristics of datasets. (a) Distribution of cognitive status categories (HC/MCI/AD) across ADReSS, NCMMSC-AD, and PUTH-AD datasets. (b) Age distribution by cognitive status categories for ADReSS and PUTH-AD datasets (NCMMSC-AD lacks age information). (c) Gender distribution by cognitive status categories across all three datasets. (d) Mini-Mental State Examination (MMSE) score distribution by cognitive status categories for the ADReSS dataset. (e-f) Montreal Cognitive Assessment (MoCA) and Hong Kong Brief Cognitive Test (HKBC) score distributions by cognitive status categories for the PUTH-AD dataset.}
    \label{fig:metadata}
\end{figure}

Figure~\ref{fig:metadata} presents the comprehensive demographic and cognitive status characteristics of all datasets. The composition shows that ADReSS contains balanced HC and AD groups with no MCI cases, while NCMMSC-AD and PUTH-AD include all three cognitive status categories. ADReSS and NCMMSC-AD exhibit relatively balanced class distributions, whereas PUTH-AD contains proportionally fewer AD cases and more MCI cases, to enhance the model's capability for detecting mild cognitive impairment. Gender distribution reveals that male participants are relatively underrepresented in the AD category for both Chinese datasets (NCMMSC-AD and PUTH-AD). Regarding age distribution, ADReSS shows that AD patients tend to be older than HC participants, while PUTH-AD demonstrates similar age distributions across cognitive status categories. For the MMSE score of ADReSS, HC participants' scores are concentrated in the 25-30 range, while AD patients exhibit a broad distribution spanning 2 to 30 points. For PUTH-AD, the MoCA score distributions show substantial separation between HC, MCI, and AD categories.

The explanation texts for training the generation component were created using Gemini-2.5-Flash, which analysed the audio recordings to produce detailed clinical rationales. This process generated 1393 explanation text samples across all datasets and tasks, with an average length of 128 words per English explanation and 562 characters for Chinese. The generated explanations focus on dementia relevant speech characteristics, including fluency patterns, semantic content, and task-specific performance indicators that inform the cognitive status classification, serving as supervision targets for training our system's explanation generation capability, ensuring that the system learns to articulate clinically meaningful features.

\subsection*{Overall Classification Performance}

\begin{figure}
    \centering
    \includegraphics[width=0.95\linewidth]{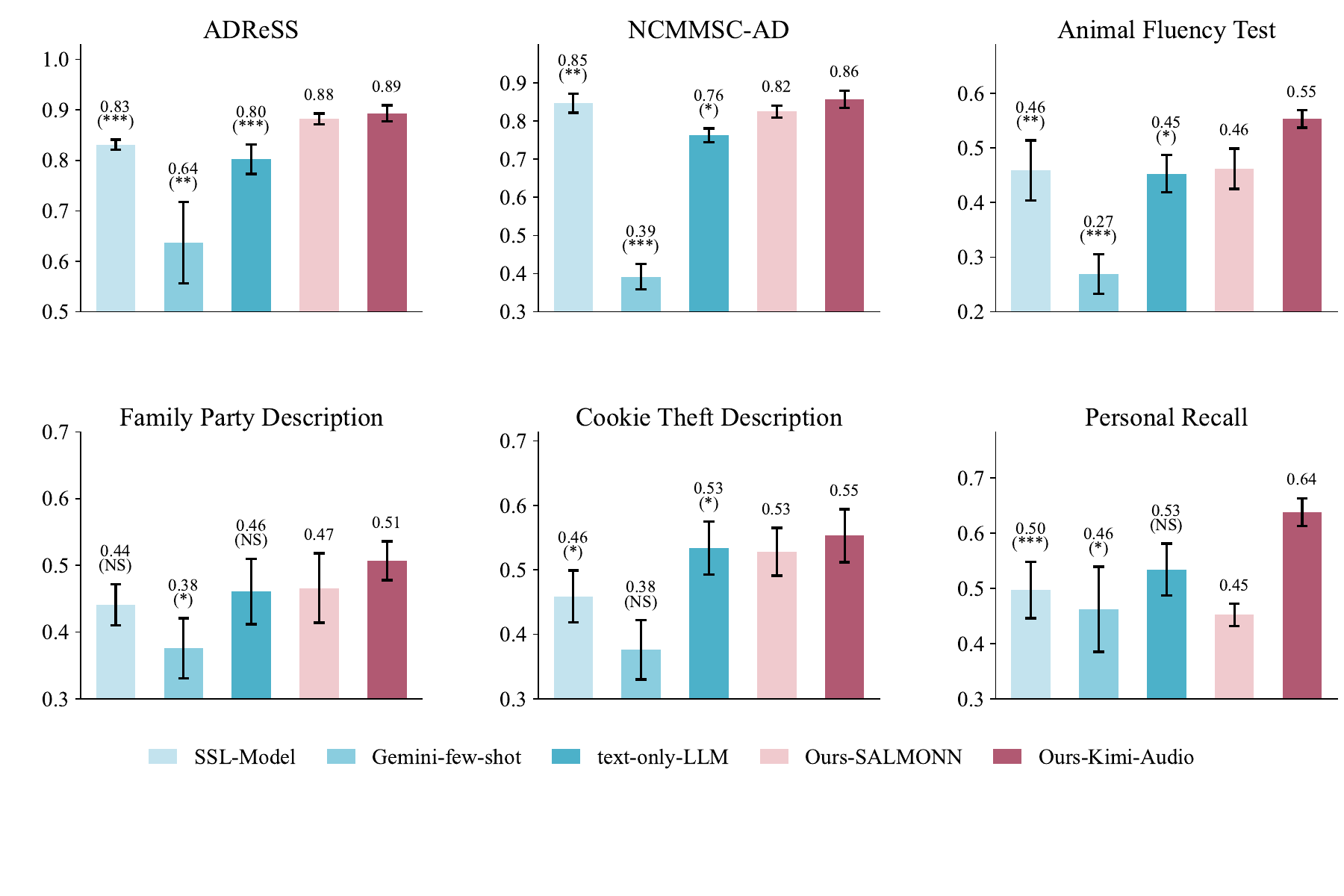}
    \caption{Classification accuracy comparison across datasets and speech tasks. Bar charts show the performance of the SSL-Model baseline, the Gemini-few-shot baseline, the Text-only-LLM baseline, and our system on six separate datasets and speech tasks. Bars show mean accuracy across five runs with different random seeds, with error bar showing standard deviation. Asterisks indicate statistical comparisons between each baseline and Ours-Kimi-Audio. Bars marked ``NS'' denote non-significant differences. $*p < 0.05; **p < 0.01; ***p < 0.001$ indicate statistical significance.}
    \label{fig:acc}
\end{figure}

\begin{table}[]
\caption{Classification performance of the Kimi-Audio model on each dataset/task. Accuracy and AUROC across five random seeds are reported in formant of mean [95\% CIs].}
\label{table: kimia_acc}
\begin{tabular}{@{}cccc@{}}
\toprule
Dataset                  & Speech Task & Accuracy             & AUROC                \\ \midrule
ADReSS                   & -           & 0.893 [0.825, 0.952] & 0.921 [0.849, 0.980] \\
NCMMSC-AD                & -           & 0.857 [0.807, 0.904] & 0.943 [0.906, 0.973] \\
\multirow{4}{*}{PUTH-AD} & AFT         & 0.553 [0.427, 0.677] & 0.675 [0.587, 0.763] \\
                         & FPD         & 0.507 [0.404, 0.612] & 0.673 [0.584, 0.763] \\
                         & CTD         & 0.553 [0.442, 0.662] & 0.681 [0.572, 0.792] \\
                         & PR          & 0.638 [0.523, 0.750] & 0.770 [0.674, 0.875] \\ \bottomrule
\end{tabular}
\end{table}

Our proposed framework employs dual output heads that simultaneously perform cognitive status classification and explanation generation, leveraging the inherent capabilities of SpeechLLMs to process raw speech and generate interpretable textual rationales. Three baselines were involved in this work for comparison: SSL-Model, an SSL model-based pipeline which combines acoustic features and linguistic features from pretrained models, Gemini-few-shot, prompting general-purpose Gemini-2.5-Flash with an identical task description and contextual examples; and text-only-LLM, employing a text LLM, Qwen3-8B for classification with transcription as input. 
Figure~\ref{fig:acc} compares classification accuracy across six dataset/task conditions, showing the mean and standard deviation across five random seeds. Complete results for all methods are provided in Supplementary Table~\ref{table: overall}.

Kimi-Audio achieved the highest mean accuracy across all six conditions, with an average of 0.666. On ADReSS, its accuracy was 0.893, compared with 0.882 for SALMONN, 0.831 for SSL-Model, 0.802 for Text-only-LLM, and 0.637 for Gemini-few-shot. On NCMMSC-AD, Kimi-Audio achieved 0.857 accuracy, slightly above SSL-Model at 0.847. Among the PUTH-AD tasks, Personal Recall yielded the highest accuracy for Kimi-Audio at 0.638. And Ours-Kimi-Audio demonstrates statistically significant improvements over baselines in 14 out of 18 pairwise comparisons. Table~\ref{table: kimia_acc} reports the accuracy and AUROC of Kimi-Audio, the best-performing model in terms of average accuracy, together with 95\% subject-level bootstrap confidence intervals (CIs). Among the four PUTH-AD tasks, Personal Recall achieved the highest mean accuracy of 0.638 (95\% subject-level bootstrap CI: [0.523, 0.750]) and AUROC of 0.770 (95\% subject-level bootstrap CI: [0.674, 0.875]), suggesting that the speech elicitation task may influence classification performance.

\subsection*{Multi-Task Learning Ablation Analysis}

\begin{figure}
    \centering
    \includegraphics[width=0.9\linewidth]{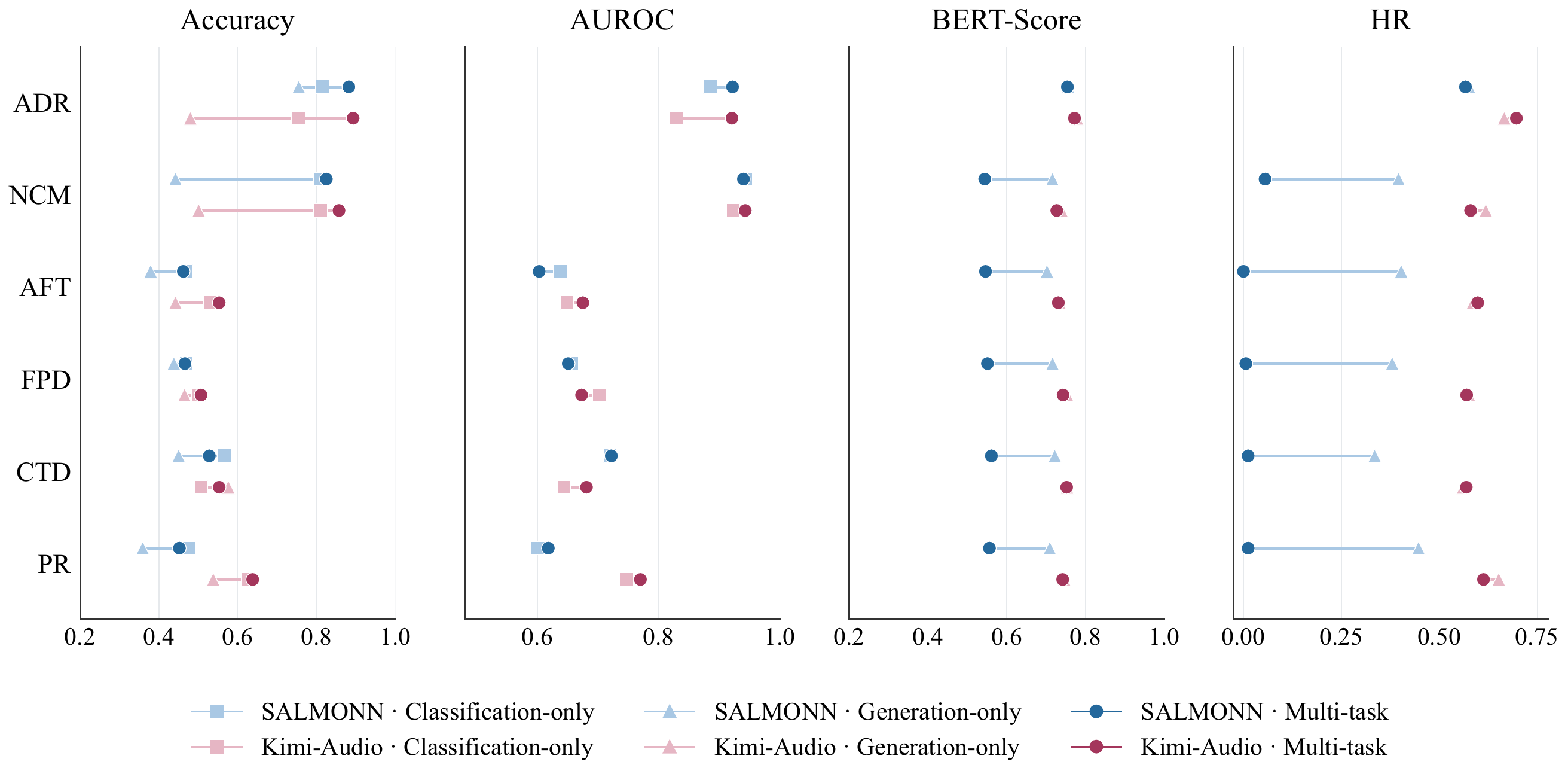}
    \caption{Ablation study comparing classification-only, generation-only, and multi-task training for SALMONN and Kimi-Audio. Panels show accuracy, AUROC, BERT-Score, and hit rate (HR), with markers indicating mean performance across five random seeds. Squares, triangles, and circles denote classification-only, generation-only, and multi-task training, respectively; lines connect training variants within each backbone. Generation-only accuracy is derived from the generated explanations. AUROC is reported for classification-only and multi-task, which have continuous class scores, and generation metrics for generation-only and multi-task which produce explanations. ADR and NCM denote ADReSS and NCMMSC-AD; AFT, FPD, CTD, and PR denote the four PUTH-AD tasks.}
    \label{fig:ablation}
\end{figure}

To assess the effectiveness of our multi-task learning framework, we conducted comprehensive ablation experiments comparing three training variants: classification-only (training solely with a cognitive status prediction objective), generation-only (training exclusively for explanation generation with classification inferred from generated text), and our proposed multi-task approach (simultaneous optimisation of both tasks). 
% The results are shown in Table~\ref{table: ablation}. 
Performance is assessed using accuracy and AUROC for cognitive status prediction, BERT-Score for semantic consistency between generated and reference explanations, and hit rate (HR) for keyword matching in generation.
Figure~\ref{fig:ablation} summarises accuracy, AUROC, BERT-Score, and HR across the six dataset/task conditions. For generation-only training, classification accuracy was derived from the generated explanations. Detailed results, including standard deviations across five runs, are provided in Supplementary Table~\ref{table: ablation}. 
Representative successful and failure examples of generated explanations presented in Supplementary Figures 1 and 2, respectively.

Compared with classification-only training, multi-task training increased Kimi-Audio’s average accuracy from 0.622 to 0.666 and average AUROC from 0.749 to 0.777. Mean accuracy increased in five conditions while maintained in FPD task, with the AUROC slightly decreased in FPD task. SALMONN showed smaller average improvements, with accuracy increasing from 0.600 to 0.602 and AUROC from 0.740 to 0.742. The largest accuracy gain occurred on ADReSS, from 0.815 to 0.882, while accuracy decreased slightly on all four PUTH-AD tasks.

Compared to the generation-only training, our multi-task approach yields a remarkable improvement in classification accuracy: SALMONN's average accuracy increases from 0.471 to 0.602, and Kimi-Audio from 0.500 to 0.666. The explanation quality largely depends on the backbone. The English-only model SALMONN shows substantial degradation on Chinese datasets (overall average BERT-Score from 0.720 to 0.585, and HR from 0.423 to 0.108). In contrast, the multilingual Kimi-Audio backbone shows little decrease in semantic similarity (average BERT-Score from 0.750 to 0.744) and keyword matching (HR from 0.610 to 0.604). The difference may come from that SALMONN is an English-only model, and jointly optimising classification and generation substantially reduced its explanation quality on Chinese data compared with generation-only training; while the multilingual model Kimi-Audio largely retained explanation quality in both languages.

\subsection*{Joint Training on Multiple Datasets Improves Generalisation}

\begin{figure}
    \centering
    \includegraphics[width=0.9\linewidth]{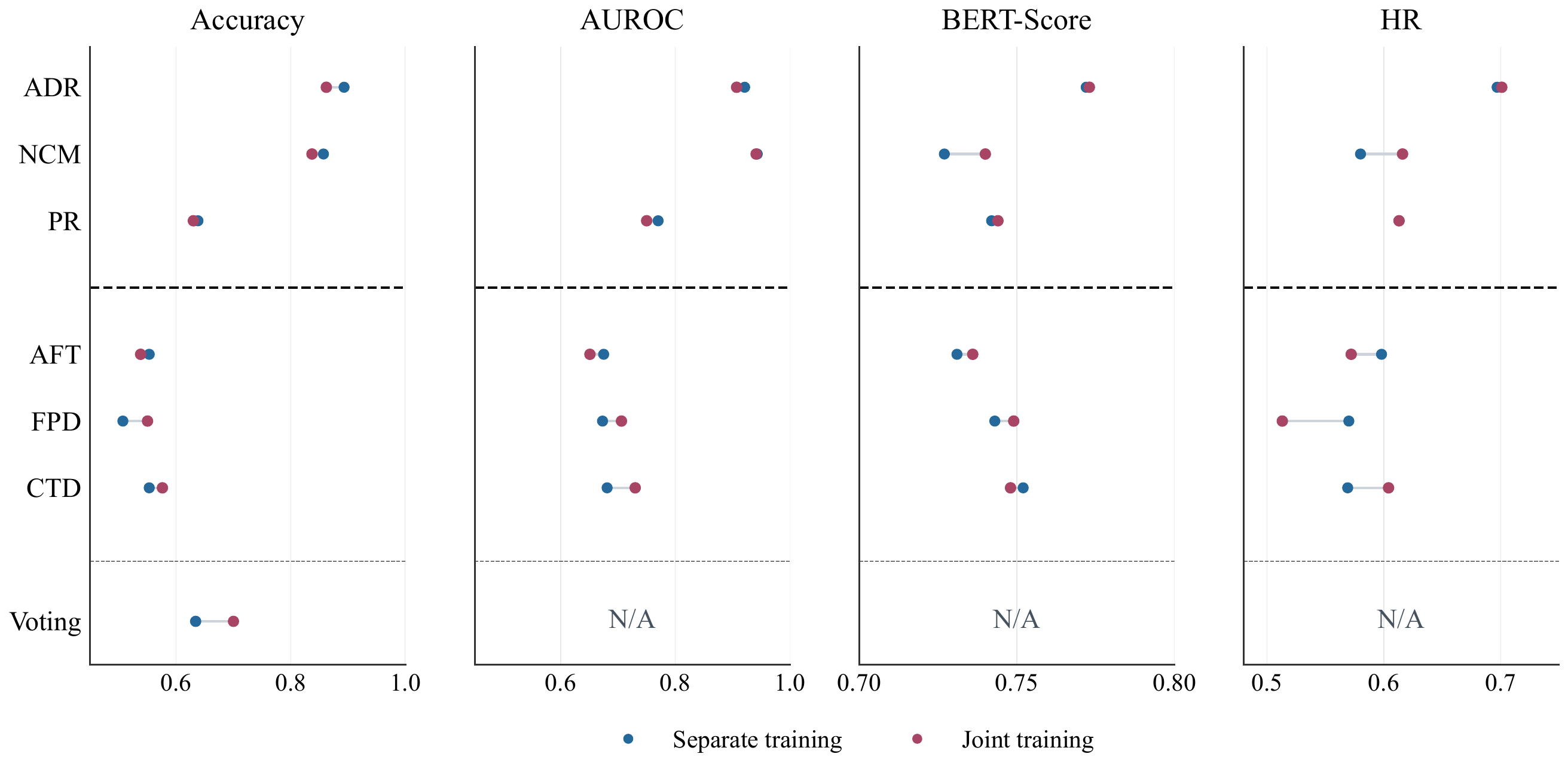}
    \caption{Comparison of separate and joint training with the Kimi-Audio backbone. Blue and red markers show mean performance across five random seeds under separate and joint training, respectively. Joint training includes ADReSS (ADR), NCMMSC-AD (NCM), and the PUTH-AD Personal Recall (PR) subset. The dashed line separates these data from the PUTH-AD AFT, FPD, and CTD subsets, which are excluded from joint training and evaluated on held-out test subjects. Lines connect the two training settings within each condition. Voting denotes participant-level hard voting across the four PUTH-AD tasks, evaluated with accuracy.}
    \label{fig:joint}
\end{figure}

To evaluate the generalisation capabilities of our approach, cross-domain joint training experiments were conducted using the Kimi-Audio backbone, which achieved superior multilingual performance in previous results. Based on Kimi-Audio's strength in processing both audio and text inputs simultaneously, the training data was enhanced with task-specific prompts to provide contextual information. The joint training dataset includes ADReSS, NCMMSC-AD, and the PR subset of PUTH-AD. The AFT, FPD, and CTD subsets of PUTH-AD were excluded from joint training and model were evaluated on held-out PUTH-AD test subjects. Results are shown in Figure~\ref{fig:joint}, with detailed results provided in Supplementary Table~\ref{tab: generalisation}.

Joint training produced mixed changes across three seen subsets. The classification performance slightly decreased, while BERT-Score and HR increased or maintained on all datasets.
Among the unseen PUTH-AD tasks, joint training increased mean accuracy on FPD from 0.507 to 0.550 and on CTD from 0.553 to 0.576. AUROC also increased on these tasks, from 0.673 to 0.706 and from 0.681 to 0.730, respectively. On AFT task, the accuracy decreased from 0.553 to 0.538. These results support that Kimi-Audio model can transfer to held-out PUTH-AD task subsets without task-specific fine-tuning. BERT-Score changed little across these subsets, while HR decreased on AFT and FPD and increased on CTD.
Under joint training, hard voting achieved an accuracy of 0.700 $\pm$ 0.029 (95\% subject-level bootstrap CI: [0.623, 0.773]), improving from 0.630 $\pm$ 0.039 for PR, the strongest individual task. Under separate training, voting achieved 0.634 $\pm$ 0.050 (95\% subject-level bootstrap CI: [0.527, 0.738]), decreased from 0.638 $\pm$ 0.025 for PR. The voting improved mean accuracy over the strongest individual task in the joint-training setting, while this improvement was not observed with separate training.

\subsection*{Class-wise Performance Analysis}
To provide a more comprehensive assessment of classification performance beyond overall accuracy, we further evaluated the joint-training Kimi-Audio model from the predefined seed-42 run using class-wise metrics.
Figure~\ref{fig:eval}(a) reports class-wise F1, sensitivity and specificity, computed in a one-vs-rest manner after pooling test predictions from ADReSS, NCMMSC-AD, and PUTH-AD voting result.
The model achieved the highest F1 score on the HC class, with F1 score, sensitivity, and specificity of 0.87, 0.88 and 0.89. The model showed relatively lower sensitivity on the MCI and AD classes, caused by the limited number of samples for MCI and AD. In contrast, the specificity for these two classes is higher. The relatively high specificity for MCI and AD suggests that the model is less likely to falsely assign non-MCI or non-AD samples to these clinically important categories.

\subsection*{Clinical Evaluation of Generated Explanations}

\begin{figure}
    \centering
    \includegraphics[width=0.95\linewidth]{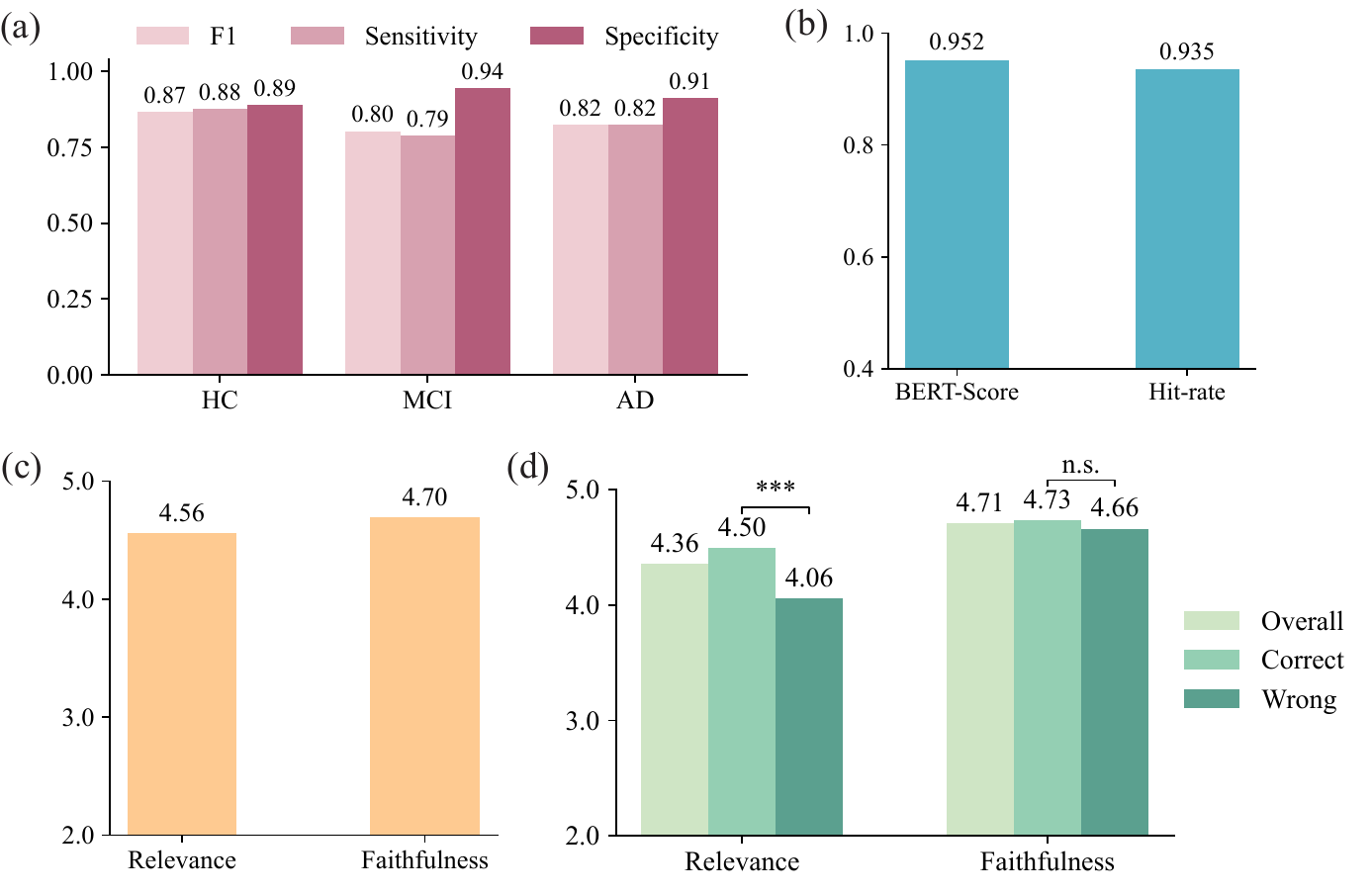}
    \caption{Clinical evaluation for classification and explanation outputs by the joint-training Kimi-Audio model using the predefined random seed of 42. (a) Class-wise F1, sensitivity, and specificity computed in a one-vs-rest manner, after pooling test predictions from ADReSS, NCMMSC-AD, and PUTH-AD voting result. (b) Agreement between original Gemini-generated explanations and clinician-corrected Gemini explanations, measured by BERT-Score and hit rate. (c) Clinician scores for Gemini-generated explanation in two dimensions, clinical relevance and evidence faithfulness. (d) Clinician scores for Kimi-Audio explanations, shown overall and separately for correctly and incorrectly classified samples. Significance was assessed by Welch’s two-sample t-test between the correct and wrong groups. $*** p < 0.001$; n.s., not significant.}
    \label{fig:eval}
\end{figure}

To evaluate the clinical quality of the generated explanations, we invited experienced clinicians to review generated explanations from the test sets of the NCMMSC-AD and PUTH-AD datasets, consisting of 327 samples. 
They scored both the Gemini-2.5-Flash supervision targets and the Kimi-Audio-generated explanations on two dimensions: clinical relevance, reflecting whether the explanation captured cognitive-decline-related speech or language features, and evidence faithfulness, reflecting whether these observations were supported by the actual recording. Clinicians also corrected the Gemini-generated explanations when needed, and the agreement between the original and corrected versions was quantified.
Before evaluation, the clinicians received training on the scoring criteria and evaluation procedure. The evaluation focused on explanation quality, and clinicians were asked to assess whether the explanation provided relevant, recording-supported evidence for the model’s stated category, without independently judging the classification itself. The ground-truth labels were not given.

To examine the quality of Gemini-generated training targets, we compared the original Gemini explanation with its clinician-revised counterparts. As shown in Figure~\ref{fig:eval}(b), the two versions exhibited high semantic agreement, achieving a BERT-Score of 0.952 and a hit rate of 0.935. These results indicate that clinician revisions introduced only limited modifications and that the key clinical content generated by Gemini was largely preserved.
Consistent with this observation, Gemini-generated explanations received high clinician ratings, with a mean clinical relevance score of 4.56 and a mean evidence faithfulness score of 4.70 (Figure~\ref{fig:eval}(c)). Together, these findings suggest that although the generated explanations do not constitute expert-written clinical rationales, they are largely consistent with clinical knowledge and supported by the underlying speech evidence, making them suitable supervision signals for explanation learning.

Figure~\ref{fig:eval}(d) reports clinician ratings for explanations generated by Kimi-Audio. The model achieved mean clinical relevance and evidence faithfulness scores of 4.36 and 4.71, respectively, indicating that the generated explanations were generally clinically meaningful and well grounded in the underlying speech evidence.
Notably, explanations associated with correctly classified samples received higher clinical relevance scores than those associated with misclassified samples, although the evaluation focused on the quality of explanations supporting a given prediction, with no groundtruth label given in evaluation. This observation suggests that explanation quality may be associated with classification correctness, highlighting the potential value of explanation outputs as an additional signal for interpreting model behaviour and identifying predictions that warrant closer clinical scrutiny.

\section*{Discussion}
\label{sec: discussion}

The variation in classification accuracy across PUTH-AD speech tasks reveals that Personal Recall achieves the highest accuracy, outperforming structured tasks, including Animal Fluency Test, Family Party Description, and Cookie Theft Description by 8.5 to 13.1 percentage points with Kimi-Audio model. This performance hierarchy likely reflects fundamental differences in task design, as Personal Recall's open-ended narrative format allows participants maximum freedom in content selection and discourse organisation, amplifying individual differences in linguistic complexity, semantic coherence, and discourse planning, where all domains are affected by cognitive decline. Unlike structured tasks that impose specific content requirements, potentially normalising speech patterns across individuals, Personal Recall enables authentic self-expression drawing upon personally meaningful experiences, providing clearer cognitive signals that our SpeechLLM models can effectively capture. 

The superior performance of our SpeechLLM-based models compared to all three baselines demonstrates the effectiveness of leveraging SpeechLLMs for cognitive impairment detection. 
The accuracy improvements over SSL baselines indicate that end-to-end audio processing with joint acoustic-semantic representations captures cognitive-status related features more effectively than traditional pipelines. 
The substantial underperformance of Gemini-few-shot, particularly on NCMMSC-AD, where it achieves only 0.392 accuracy (while our model achieved accuracy of 0.857), reveals that despite sophisticated reasoning capabilities, general-purpose language models lack sufficient domain-specific understanding to distinguish subtle differences between HC, MCI, and AD categories without targeted fine-tuning. 
The text-only-LLM baseline achieved competitive performance on the PUTH-AD Cookie Theft Description data. However, its average accuracy on all datasets remained close to the SSL baseline and lower than Ours-Kimi-Audio. This suggests that transcript-based LLMs can capture useful linguistic cues, but still lose acoustic and prosodic information that may be important for cognitive decline detection. In contrast, the SpeechLLM-based Kimi-Audio model directly processes raw speech and jointly models acoustic and semantic information, providing a more comprehensive basis for classification.

The ablation analysis reveals that multi-task learning of classification and explanation generation objectives generally maintains or improves classification accuracy compared to classification-only training, with improvements particularly pronounced on ADReSS dataset.
The gains from multi-task training may reflect a regularisation effect of explanation supervision, which encourages the model to attend to speech characteristics relevant to cognitive status.
The explanation generation objective encourages the model to identify specific acoustic and linguistic features that are relevant to cognitive impairment, strengthening latent representations and making them more discriminative for classification. 
However, the generation-only approach's poor classification performance demonstrates that explanation generation alone cannot substitute for explicit classification objectives, as fluent explanations often lack the precision supervision for reliable diagnostic inference. This finding underscores that interpretability components should complement rather than replace dedicated classification heads in clinical decision support systems, with our multi-task approach balancing classification performance and explanation generation.

The cross-domain joint training experiments provide compelling evidence for generalisability across task types. Remarkably, the Family Party Description task, completely absent from training data, achieves higher classification accuracy than under separate training. This may reflect that the model has learned generalisable patterns associated with cognitive impairment rather than task-specific artefacts, by recognising that picture description tasks require similar cognitive processes whose disruption manifests in comparable speech patterns regardless of specific content. 
The improved BERT-Scores across datasets, including unseen data, suggest that exposure to diverse training examples enhances the model's ability to articulate reasoning in linguistically varied and contextually appropriate ways. The ensemble voting approach across all PUTH-AD tasks shows substantial improvement from 0.634 to 0.700 accuracy with joint training, indicating that the generalisation brought by joint training provides more reliable overall assessment. This cross-task generalisation is crucial for clinical deployment where assessment protocols vary across institutions and cultures, suggesting that a single model architecture could serve diverse clinical contexts while maintaining classification reliability.

These findings have important implications for developing real-world cognitive screening tools. The model's explanation generation capability addresses a critical barrier to AI adoption by providing natural language rationales that clinicians can evaluate and verify, enabling workflows where AI serves as an intelligent assistant. The generalisation to held-out speech tasks suggests that the SpeechLLM-based architecture could serve diverse clinical contexts with different speech tasks, significantly enhancing its practical clinical value. 

This study has several limitations.
The relatively limited sample sizes across datasets, particularly for MCI and AD classification, may limit statistical power and generalisation to broader populations.
The PUTH-AD diagnostic labels were operationally defined using cognitive screening score thresholds (MoCA and HKBC), introducing potential diagnostic-label noise.
Differences in age and sex distributions across cognitive status groups may also have influenced classification performance.
Although cross-task generalisation experiments were conducted, all datasets were used during model development, and no fully independent external cohort was used for validation.
The explanation generation task relies on Gemini-2.5-Flash-generated explanations as training targets. Although clinician scoring and correction suggested that these targets were largely clinically relevant and consistent with the speech evidence, LLM-generated explanations are not equivalent to fully human-written clinical rationales. They may still contain unsupported observations, overconfident statements, or teacher-model biases, which the trained model can inherit.
The future work should focus on expanding training data to include additional languages, cultural contexts, and clinical populations to further validate and enhance the model's generalisation capabilities. 
Furthermore, domain experts should be involved to evaluate the system in closed-loop clinical workflows, thereby establishing validated protocols for human-AI collaboration in cognitive assessment.

\section*{Methods}
\label{sec: methods}

\subsection*{Datasets and Speech Tasks}

Our primary dataset, \textit{PUTH-AD}, was collected at Peking University Third Hospital and Peking University Sixth Hospital, and consists of 166 participants, each completing four distinct speech assessment tasks designed to capture different aspects of cognitive and linguistic functioning relevant to Alzheimer's disease detection:
\begin{itemize}
    \item \textit{Animal Fluency Test}: This task utilised the \textit{Animal Fluency Test}, traditionally a measure of cognitive executive function and semantic memory access~\cite{whiteside2016verbal}. Participants were asked to list as many animals as possible, revealing both cognitive processing speed and linguistic retrieval abilities that are commonly impaired in AD and MCI.
    \item \textit{Cookie Theft Description}: Participants were asked to describe the classic Cookie Theft picture, a standardised assessment tool widely used in aphasia and dementia evaluation for its ability to elicit narrative discourse and reveal missing details, pronoun confusion and word-finding difficulties~\cite{cummings2019describing}. Given that the data collection was conducted in Chinese, the original image was modified by replacing the English word ``cookie'' with the Chinese equivalent while maintaining all other visual details unchanged (see Figure~\ref{fig:overall}).
    \item \textit{Family Party Description}: We created an original picture depicting a Chinese New Year family dinner scene (see Figure~\ref{fig:overall}), and participants were asked to describe this picture. Unlike the Cookie Theft picture, the family party one is designed to be more culturally relevant and familiar to Chinese elderly participants.
    \item \textit{Personal Recall}: Participants were given three minutes to provide a brief introduction about their family situation, hobbies, and health status. This free-form narrative task evaluates spontaneous speech production, autobiographical memory access, and discourse organisation, all of which can be affected by cognitive decline.
\end{itemize}

An Android application was developed for speech data collection, which runs on mobile devices and utilises the phone's built-in microphone to simulate real-world application scenarios.
The application used for data collection guided participants through six sequential steps: (1) Task instructions and rules are presented with audio playback; (2) Cookie Theft picture description task; (3) Animal fluency task where participants list items from a given category; (4) Personal recall task where participants share family situations and personal experiences;  (5) Family Party picture description task depicting a Chinese New Year dinner scene; (6) Completion confirmation. Each task begins with written instructions followed by a recording button that participants press to start and stop their response.

Eligibility screening considered relevant medical history and medication use, including psychiatric and neurological conditions, severe systemic disease, and alcohol or substance misuse. Participants were aged 40–80 years and were required to understand and cooperate with the testing procedure, with sufficient vision, hearing, verbal communication, and hand function to complete the tasks and operate the recording device.
All recordings were conducted in quiet, soundproof hospital environments or community facilities.
All physicians involved in data collection and in the manual evaluation were neurologists from the Department of Neurology, Peking University Third Hospital. They have clinical experience in the diagnosis and management of Alzheimer's disease and cognitive disorders, are engaged in both clinical and basic research in neurology, and received standardized training before performing the study procedures.
All procedures involving human participants were conducted in accordance with the Declaration of Helsinki. The PUTH-AD data-collection protocol was approved by the Medical Ethics Committee of Peking University Sixth Hospital (Institute of Mental Health; approval no. 2024 [No. 20]) and the Peking University Third Hospital Medical Science Research Ethics Committee (approval no. IRB000006761-M2024136). Written informed consent was obtained from all participants before data collection.

Participants underwent cognitive assessment using two validated scales: the Montreal Cognitive Assessment (MoCA)~\cite{nasreddine2005montreal} and the Hong Kong Brief Cognitive Test (HKBC)~\cite{chiu2018development, sun2022validation}. Cognitive status was operationally defined based on score thresholds: MoCA scores of 26–30 indicated normal cognition, 18–25 indicated MCI, and $\le$ 17 indicated dementia-level impairment; HKBC scores of 21–30 indicated normal cognition, 17–20 indicated MCI, and $\le$ 16 indicated dementia-level impairment. 
It should be noted that these score-based classifications reflect cognitive screening outcomes rather than formal clinical diagnoses of AD, as no independent neurological adjudication or biomarker confirmation was performed. In cases where the two scales yielded discordant classifications, the more impaired classification was assigned. The clinicians who supervised data collection did not participate in model development or evaluation. The labels HC, MCI, and AD denote these three score-defined categories, respectively, with AD used as the dataset label for dementia-level impairment.

To enhance model generalisability, several established open-source datasets, covering both English and Mandarin Chinese, were collected into our study.

The ADReSS~\cite{luz2020alzheimer} and ADReSSo~\cite{luz2021detecting} challenges were organised at Interspeech 2020 and Interspeech 2021 respectively, with most data selected from the Pitt corpus~\cite{becker1994natural} from the DementiaBank database. We removed duplicates and merged the data from both challenges to form the \textit{ADReSS dataset}, comprising 330 participants who each provided a recording describing the ``Cookie Theft'' picture in English, with binary diagnostic labels (AD/HC).

The 
{NCMMSC-AD dataset} is a Mandarin speech dataset from a challenge held at the NCMMSC 2021 conference. This dataset includes 399 audio segments from 175 participants performing various cognitive tasks, including verbal fluency, lexical retrieval (antonyms/synonyms), picture description, self-introduction, \textit{etc.} Each sample is labelled as HC, MCI, or AD, providing three-class classification data. This diverse-task dataset helps the model learn a range of speech manifestations of cognitive impairment. The NCMMSC-AD recordings had undergone pause removal as part of the original dataset preprocessing.

The ADReSS, ADReSSo, and NCMMSC-AD datasets comprised previously collected and de-identified data. Their use in the present study complied with the respective data-access requirements and the ethical approvals and informed-consent procedures reported by the original data providers. No additional participant recruitment or intervention was conducted for these secondary analyses.

Gemini-2.5-Flash was used to construct explanation supervision because it supports direct audio input and explanation generation in both English and Chinese. The suitability of these targets was assessed through clinician ratings and comparison with clinician-revised explanations.
Gemini was provided with the recording, task information, and the dataset label to generate an explanation for the given label.
The prompt structure consisted of three components: i) a clear task definition explaining the objective of determining whether the speaker exhibits characteristics of AD, MCI, or HC; ii) a comprehensive description of typical speech manifestations observed in AD patients; iii) a brief explanation of the specific speech task being performed by the participant. 
To ensure thorough analysis of the speech content, we implemented a two-stage generation pipeline.
In the first stage, the model generated a detailed analysis of acoustic and linguistic characteristics, including lexical diversity, repetition, word-finding difficulty, fluency, semantic coherence, and discourse organisation.
In the second stage, the model was instructed to synthesise and simplify the detailed analysis from the first stage, retaining only the most salient indicators to produce concise, clinically relevant explanations. Both the prompt language and output text language were matched to the audio input language to maintain consistency. A representative prompt template is shown in Supplementary Figure 3.

\subsection*{Model Architecture}

SpeechLLMs represent a significant advancement in multimodal AI, extending the capabilities of large language models by incorporating audio understanding while maintaining their text processing and prompt-following abilities~\cite{tang2023salmonn, rubenstein2023audiopalm}. These models are built upon pre-trained text language models and enhanced with audio comprehension capabilities, enabling them to process both audio and textual inputs simultaneously and making them particularly suitable for multi-linguistic and multi-speech-task applications. The architecture typically employs a two-stage approach~\cite{peng2024survey}: in the first stage, raw audio waveforms are converted into either continuous embeddings or discrete tokens, which are then concatenated with textual input tokens to form a unified multimodal representation. In the second stage, a Transformer-based model processes this combined audio-text sequence to understand the multimodal content and generate appropriate outputs. The key advantage of SpeechLLMs extends beyond eliminating the need for automatic speech recognition; most importantly, they possess sophisticated audio understanding and reasoning capabilities that enable them to analyse complex acoustic patterns and generate explanatory natural language text that articulates the underlying reasoning process~\cite{peng2024survey, ding2025kimi}, making them ideally suited for interpretable medical diagnosis tasks.

\begin{figure}
    \centering
    \includegraphics[width=0.9\linewidth]{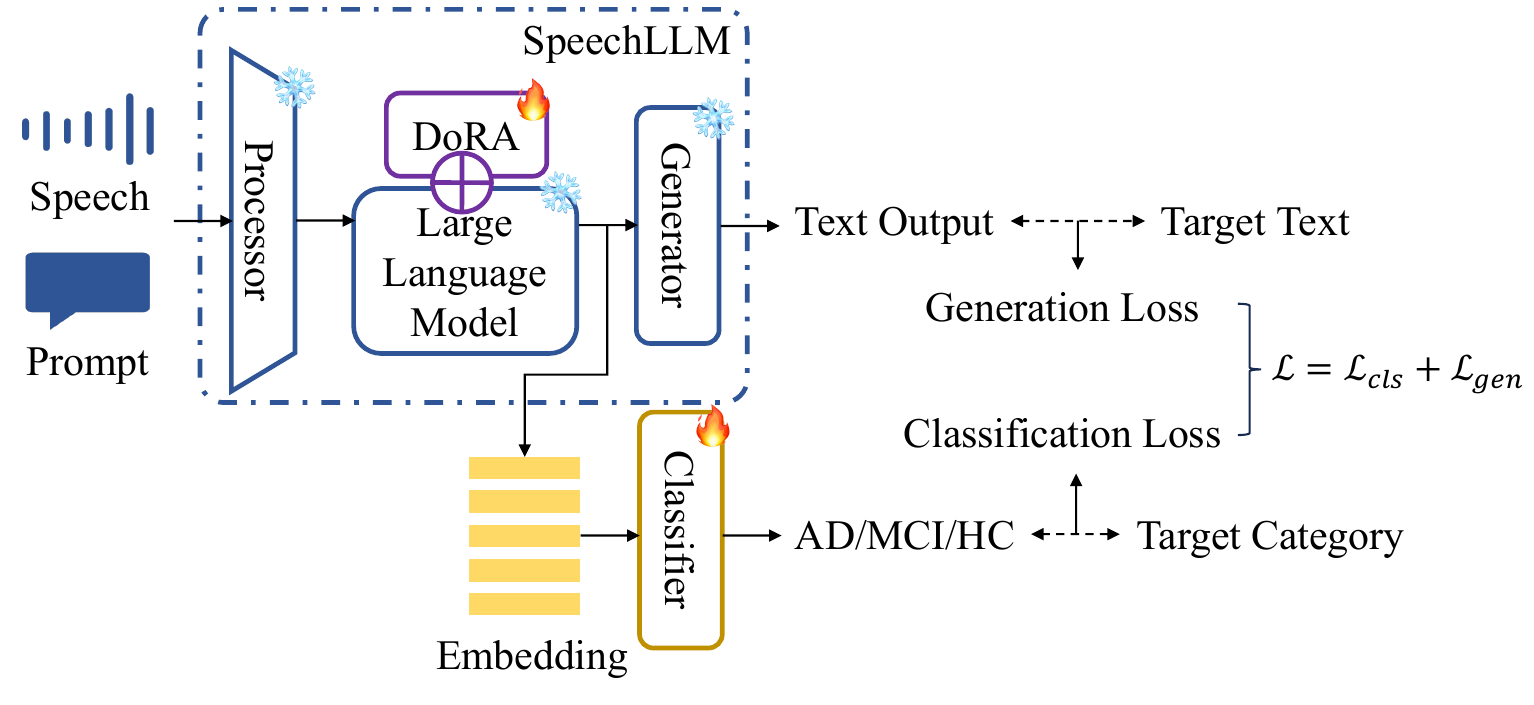}
    \caption{Model Architecture. Dual output heads simultaneously do classification and generation. The model is optimised through both losses.}
    \label{fig:model}
\end{figure}

Building upon this, we proposed a unified multi-task framework that extends the standard SpeechLLM architecture with dual output heads to simultaneously perform cognitive status classification and explanation generation, as shown in Figure~\ref{fig:overall}. Model details are shown in Figure~\ref{fig:model}. 
The classification head takes the hidden state corresponding to the last input token from the final Transformer layer and applies a linear projection to obtain class logits, followed by Softmax to produces probabilities.
In parallel, the generation head leverages the model's inherent text generation capabilities to produce natural language explanations that justify the classification decision. The classification component is optimised using standard cross-entropy loss, while the generation component is trained via a teacher-forcing approach. The joint training strategy optimises both objectives simultaneously through a combination of classification and generation losses ($\mathcal{L}=\mathcal{L}_\text{cls} + \mathcal{L}_\text{gen}$), encouraging the model to learn representations that are both discriminative for classification and interpretable through natural language descriptions.

Our approach is implemented and evaluated using two 7-billion-parameter SpeechLLMs: SALMONN~\cite{tang2023salmonn} and Kimi-Audio~\cite{ding2025kimi}. SALMONN is an English SpeechLLM enabling speech, audio events, and music inputs. Kimi-Audio is a multilingual model pretrained on over 13 million hours of diverse audio data, capable of audio understanding, generation, and conversation. To enable targeted learning of AD-related knowledge, the DoRA (Weight-Decomposed Low-Rank Adaptation)~\cite{liu2024dora} module is applied to the query and key matrices within the language model backbone's attention layers, allowing the model to adapt its attention mechanisms specifically for cognitive impairment detection while maintaining computational efficiency. 
Only the DoRA parameters and the classification head are updated, while the remaining pretrained parameters are frozen, resulting in 4.4 million trainable parameters for SALMONN and 3.2 million for Kimi-Audio.

\subsection*{Baseline Models}

\textit{SSL-Model}: Self-supervised learning (SSL) models acquire rich, generalisable feature representations from large-scale data, and, when fine-tuned for specific downstream tasks, can develop specialised capabilities. This approach has shown promising results in previous AD detection studies~\cite{koo2020exploiting, zhu2021wavbert, chen2023cross, cui2023transferring}. We implement a traditional pipeline approach using SSL models as feature extractors combined with downstream classification heads. This baseline extracts both acoustic and linguistic features by employing separate speech and text SSL models, then concatenates these features before feeding them into a downstream classifier for cognitive status prediction. Whisper-Large-v3~\cite{radford2023robust} model was applied for ASR. For English data, WavLM-Base-Plus~\cite{chen2022wavlm} was utilised for acoustic feature extraction, and BERT-base-uncased~\cite{kenton2019bert} for linguistic feature from ASR transcripts. For Chinese data, Wav2Vec2-large-xlsr-53-chinese-zh-cn~\cite{grosman2021xlsr53-large-chinese} (an XLSR model~\cite{conneau2021unsupervised} fine-tuned on Chinese speech data) was applied for acoustic features and BERT-base-Chinese for linguistic features. The specific model architecture follows the implementation described in~\cite{cui2023transferring}.

\textit{Gemini-few-shot}: Gemini-2.5-Flash, which serves as the explanation text generation model, is directly evaluated on the classification task to assess how well general-purpose LLMs perform without domain-specific finetuning. The input prompt maintains identical task description and formatting as those used during the explanation text generation process, with the prompt template shown in Supplementary Figure 4. A few-shot learning approach was adopted by providing the Gemini model with two random examples for each category from the training set as contextual demonstrations, followed by the test audio for prediction. 

\textit{Text-only-LLM}: To determine whether text-only LLM fine-tuning on ASR transcripts can achieve comparable performance to direct raw-speech modelling, we implemented a text-only-LLM baseline. Raw speech recordings were first transcribed using the Whisper model, which were then used as text input to Qwen3-8B~\cite{yang2025qwen3} for cognitive status classification. To ensure a fair comparison with the Kimi-Audio classification setting, the text-only-LLM baseline used the same classification head design, DoRA fine-tuning strategy, set of five random seeds, and train/development/test splits.

\subsection*{Statistical Analysis}

All dataset evaluations followed a cross-subject train/development/test paradigm to ensure proper generalisation assessment. 
For the ADReSS dataset, the official test set from the ADReSSo challenge was used as the test split, and likewise, the NCMMSC-AD dataset evaluation also employed the official test partition provided by the challenge organisers.
For our PUTH-AD dataset, 52 subjects were selected as the test set, while ensuring balanced representation across cognitive status categories. From the remaining training data in each dataset, approximately 20\% was randomly sampled to form validation sets for hyperparameter tuning and model selection.
The same subject-level train/development/test split was used for all four PUTH-AD tasks, ensuring that recordings from test subjects were never included in training, regardless of task type.
Each trainable model configuration was evaluated in five runs using five different random seeds. The subject-level training, development, and test partitions were held fixed across runs. Class-wise analysis and clinician evaluation used the joint-training Kimi-Audio model from the predefined seed-42 run.
For the Gemini-few-shot baseline, five runs were performed using different random seeds to sample the few-shot examples from the training set, with two examples per class in each run.

Classification performance was evaluated using accuracy and the area under the receiver operating characteristic curve (AUROC). Metrics were computed separately for each run and summarised as the mean $\pm$ standard deviation across five runs.
Accuracy was calculated from the predicted class labels, whereas AUROC was calculated from continuous class prediction scores. For ADReSS, binary AUROC was computed with AD as the positive class. For NCMMSC-AD and PUTH-AD, multiclass AUROC was computed using weighted-averaged one-vs-rest AUROC.
AUROC was not reported for the Gemini-few-shot baseline or the generation-only ablation because these settings produced categorical predictions without continuous class probabilities.
For the explanation generation task, we employed two complementary evaluation metrics: BERT-Score and hit rate (HR). BERT-Score~\cite{zhang2020bertscore} leverages pre-trained contextual embeddings from BERT to match words in candidate and reference sentences through cosine similarity, providing a measure of semantic consistency between model-generated explanations and target texts. To accommodate both Chinese and English text evaluation, the BERT-base-multilingual-cased model was utilised as the scoring model. Hit rate evaluates the coverage of clinically relevant keywords in the generated explanations by extracting keywords from both generated and target texts using GPT-4o-mini, then calculating the recall of these critical terms in the model outputs. For the generation-only ablation study presented in Table~\ref{table: ablation}, where no direct classification output was available, we employed GPT-4o-mini to analyse the generated explanation texts and infer the diagnostic category, using this derived classification for accuracy computation. 
Subject-level bootstrap 95\% CIs were also calculated. For each dataset/task condition, 95\% CIs were estimated using 10,000 bootstrap resamples of test subjects with replacement. Accuracy used ordinary subject resampling, whereas AUROC used resampling within each ground-truth class, preserving the original class counts. Each selected subject’s predictions from all five runs were retained together. Within each resample, metrics were calculated separately for each run and then averaged across the five runs. The 2.5th and 97.5th percentiles of the resulting distribution defined the CI.
Statistical comparisons used predictions from the predefined seed-42 runs. Comparisons between Kimi-Audio and the SSL-Model and Text-only-LLM baselines used Wilcoxon signed-rank tests on paired probabilities assigned to the ground-truth class. The comparison with Gemini-few-shot used an McNemar’s test, implemented as a binomial test on samples for which only one model predicted correctly, under the null hypothesis that either model was equally likely to be correct.

For each PUTH-AD test participant, predictions from the four speech tasks were combined by hard voting, with the class receiving the most votes selected as the final prediction. Ties were resolved by selecting the tied class that appeared first in a prespecified task order: AFT, FPD, CTD, and PR. Voting accuracy was computed separately for each random seed and summarised as the mean $\pm$ standard deviation across five runs, with a corresponding 95\% CIs.

\subsection*{Implementation Details}

All experiments were conducted on a single NVIDIA A800 GPU. We employed DoRA for parameter-efficient fine-tuning on the LLM backbone, with LoRA rank set to 8, LoRA alpha set to 32, and LoRA dropout set to 0.1. This configuration resulted in 4.4 million trainable parameters for SALMONN and 3.2 million trainable parameters for Kimi-Audio. 
SALMONN and Kimi-Audio shared the same optimisation settings: AdamW with an initial learning rate of $1\times10^{-4}$, a cosine learning-rate scheduler, and a weight decay of 0.005. Both models used an effective batch size of 4. SALMONN used a mini-batch size of 4 without gradient accumulation, whereas Kimi-Audio used a mini-batch size of 1 with four gradient accumulation steps.
For the SpeechLLM experiments, models were trained for 10 epochs in each run. For classification-only and multi-task training, the checkpoint with the highest development-set classification accuracy was selected. For generation-only training, the checkpoint with the lowest development-set generation loss was selected. The selected checkpoints were then evaluated on the test sets.
Results are reported as the mean $\pm$ standard deviation across the five runs, together with 95\% CIs.
The Kimi-Audio inference required approximately 7 seconds per sample and 34 GB of GPU memory, favouring server-based deployment.

The SSL-Model baseline followed the implementation and training protocol of our previous work~\cite{cui2023transferring}, including its data augmentation procedure. It used a batch size of 128 and the Adam optimiser, with a linear learning-rate schedule decreasing from $4\times10^{-5}$ to $1\times10^{-5}$. Its downstream classifier comprised two Transformer encoder layers with a dropout rate of 0.2.  The Text-only-LLM baseline used the same optimisation settings and effective batch size as Kimi-Audio.

The classification head takes the hidden state corresponding to the last input token from the final Transformer layer and applies a linear projection to obtain class logits. Under separate training, the head outputs two classes for ADReSS and three classes for NCMMSC-AD and PUTH-AD. Under joint training, a shared three-class head is used, with HC, MCI, and AD encoded as 0, 1, and 2, respectively. ADReSS contributes samples with labels 0 and 2 only. During evaluation on ADReSS, only the HC and AD logits are retained for computing binary classification accuracy and AUROC, with AD treated as the positive class.

\subsection*{Clinical Evaluation}

For clinician evaluation, Kimi-Audio explanations were generated by the joint-training model from the run with the predefined random seed of 42, which was specified in advance. We evaluated the clinical quality of generated explanations of 327 test samples from the NCMMSC-AD and PUTH-AD datasets, with no sample excluded in clinical evaluation. These two datasets were selected because both the speech recordings and explanations are in Chinese, allowing reliable assessment by native Chinese-speaking clinicians. Five experienced neurologists were invited to review the explanations. 
The evaluation consisted of two procedures. Firstly, explanations were scored on two dimensions: clinical relevance and evidence faithfulness. Clinical relevance measured whether the explanation identified speech or language characteristics associated with cognitive decline. Evidence faithfulness assessed whether the statements in an explanation were supported by the actual speech recording, with lower scores indicating more unsupported or hallucinated content. 
Second, clinicians revised the Gemini-generated explanations where necessary. The agreement between the original Gemini-generated explanations and the clinician-revised versions was then quantified using BERT-Score and hit rate. 
Each sample was assessed by one neurologist, who rated both the Gemini-generated and Kimi-Audio-generated explanations for clinical relevance and evidence faithfulness on a scale from 1 to 5, with higher scores indicating better quality.

To facilitate annotation, we developed a web-based evaluation interface. A screenshot of the interface is shown in Supplementary Figure 5. For each sample, the left panel provides the speech recording, which can be played directly in the browser, together with the corresponding transcription and task instruction. For picture description tasks, the task image is also displayed to provide the full task context. The right panel contains the annotation fields for explanation evaluation. From top to bottom, it shows the original Gemini-generated explanation, the clinician scoring fields for the Gemini explanation, an editable text box for correcting the Gemini explanation, the Kimi-Audio-generated explanation, and the clinician scoring fields for the Kimi-Audio explanation. For both Gemini and Kimi-Audio explanations, clinicians score two dimensions: relevance and faithfulness.
To make the annotation process easier and improve scoring reliability, all interface text, task information, transcriptions, generated explanations, and scoring instructions are presented in Chinese. The evaluation program is deployed on a local server to avoid data leakage.

\backmatter

\section*{Data Availability}

The ADReSS and ADReSSo datasets analyzed during the current study are publicly available through TalkBank. The NCMMSC-AD and PUTH-AD datasets are not publicly available due to ethical restrictions but are available from the corresponding author on reasonable request, subject to the respective data use and ethical approval requirements.

\section*{Code Availability}
The code is publicly available at \url{https://github.com/cuiziyun/kimia-explainable-AD} under the Apache-2.0 license. The repository includes the complete environment specification, software versions, configuration files, prompt templates, and key parameters required to reproduce the analyses reported in this study.

\section*{Acknowledgements}
Not applicable.

\section*{Funding}
This work was supported by the OPPO Research Fund and the National Natural Science Foundation of China (Grant No. 62501336). OPPO, as a funder, participated in the study design and manuscript preparation. The National Natural Science Foundation of China had no role in the study design, data analysis, manuscript preparation, or decision to publish.

\section*{Author Contributions}
Z.C., W.W., X.G, Y.Z., N.L., and C.Z. designed research; Z.C., C.S., and S.Y. performed research; W.Z. and J.W. provided datasets; Z.C., W.W., and C.Z. analysed data; Q.Y., H.Z., Y.L., and N.L. contributed to clinical validation; and Z.C., W.W., X.G. and C.Z. wrote the paper. All authors read and approved the final manuscript.

\section*{Competing Interests}
X.G., Y.Z., and Y.L. are employees of OPPO Research Institute. The other authors do not have a competing interest.

\bibliography{sn-bibliography} % common bib file

\clearpage

\setcounter{figure}{0}

\renewcommand{\figurename}{Supplementary Figure}

\renewcommand{\thefigure}{\arabic{figure}}

\setcounter{table}{0}
\renewcommand{\tablename}{Supplementary Table}
\renewcommand{\thetable}{\arabic{table}}

\section*{Supplementary Information}

\subsection*{Generation Examples}
\label{supp:generate-examples}

\begin{figure}[h]
    \centering
    \begin{subfigure}{\linewidth}
        \centering
        \includegraphics[width=1.0\linewidth]{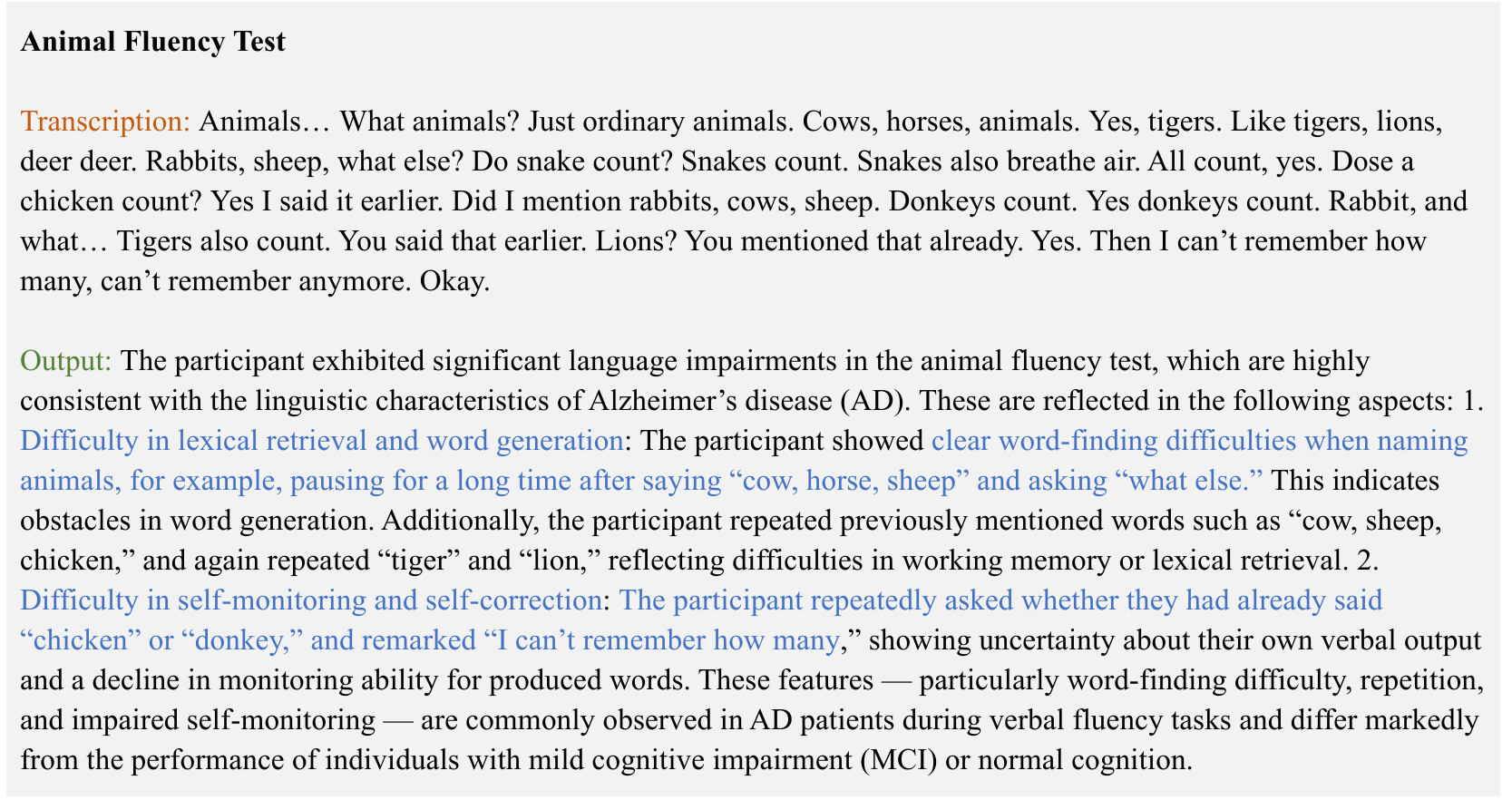}
        \caption{A successful example on the Animal Fluency Test task. The cognitive status label is AD, and the model correctly predicts AD. The blue text-marked phrases indicate the key cognitive-related evidence identified by the model, including lexical retrieval difficulty, word-generation difficulty, repeated item production, and impaired self-monitoring. These features are reflected in the participant’s pauses, repeated animal names, uncertainty about whether certain animals had already been mentioned, and inability to recall more items.}
    \end{subfigure}
    \begin{subfigure}{\linewidth}
        \centering
        \includegraphics[width=1.0\linewidth]{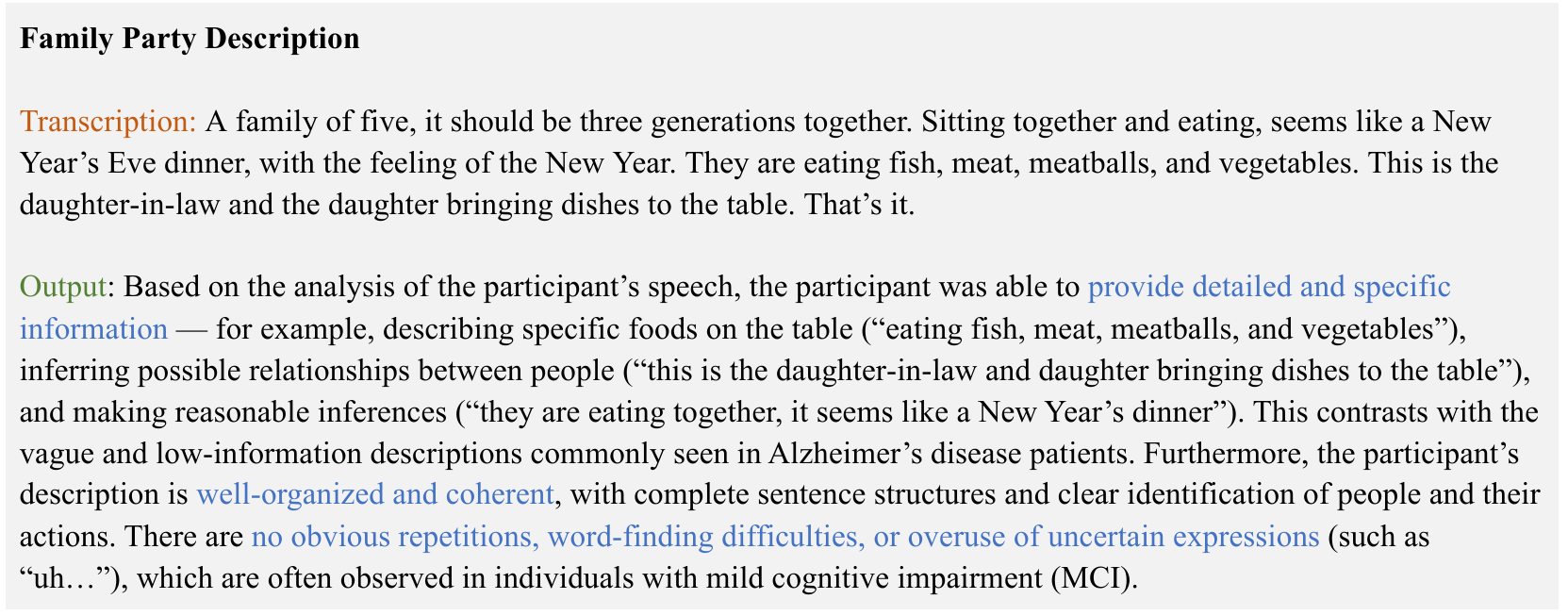}
        \caption{A successful example of the Family Party Description task. The cognitive status label is HC, and the model correctly predicts HC. The blue text-marked phrases indicate the key evidence supporting the model’s prediction, including detailed and specific information, coherent organization, clear description of people and actions, and the absence of obvious repetition, word-finding difficulty, or excessive uncertain expressions. These features suggest preserved language ability and intact scene-description performance.}
    \end{subfigure}
    \caption{Two representative successful examples of generated explanations on the AFT task and the FPD task. Each example includes the ASR transcription of the participant's speech, the system's generated explanation output, and the ground truth category. The original transcription and output are in Chinese, and translated to English here.}
    \label{fig:success}
\end{figure}

Supplementary Figure~\ref{fig:success} presents two successful examples of explanation generation, illustrating how the model articulates reasoning based on observed speech characteristics. For the AFT task, the system identified key indicators of cognitive impairment in an AD patient's performance, including difficulty in lexical retrieval and word generation, repetition, and difficulty in self-monitoring. In the FPD task, the system recognises a healthy control participant's ability to provide detailed and specific information, describing foods, family relationships, and vegetable names with clarity. The explanation highlights the coherent discourse structure, well-organised narrative with complete sentence structures, and clear delineation of people and actions, noting the absence of repetition, word-finding difficulties, or uncertain expressions that would suggest cognitive impairment.

\begin{figure}[h]
    \centering
    \begin{subfigure}{\linewidth}
        \centering
        \includegraphics[width=1.0\linewidth]{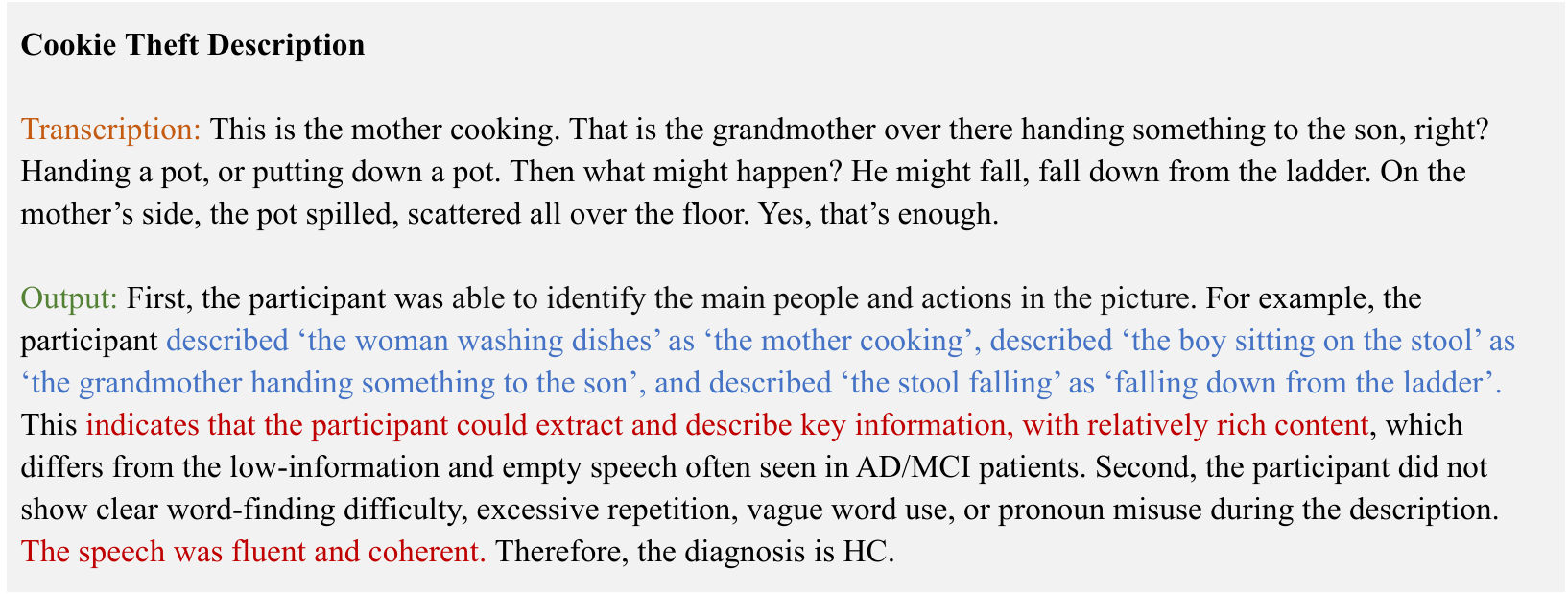}
        \caption{Failure case on the Cookie Theft Description task. The cognitive status label is AD, but the model incorrectly predicts HC. The blue text marks inaccurate or mismatched descriptions of picture elements. The red text highlights the model’s erroneous rationale: it treats these inaccurate descriptions as evidence of rich content, fluency, and preserved language ability, thereby overestimating the participant’s cognitive-linguistic function.}
    \end{subfigure}
    \begin{subfigure}{\linewidth}
        \centering
        \includegraphics[width=1.0\linewidth]{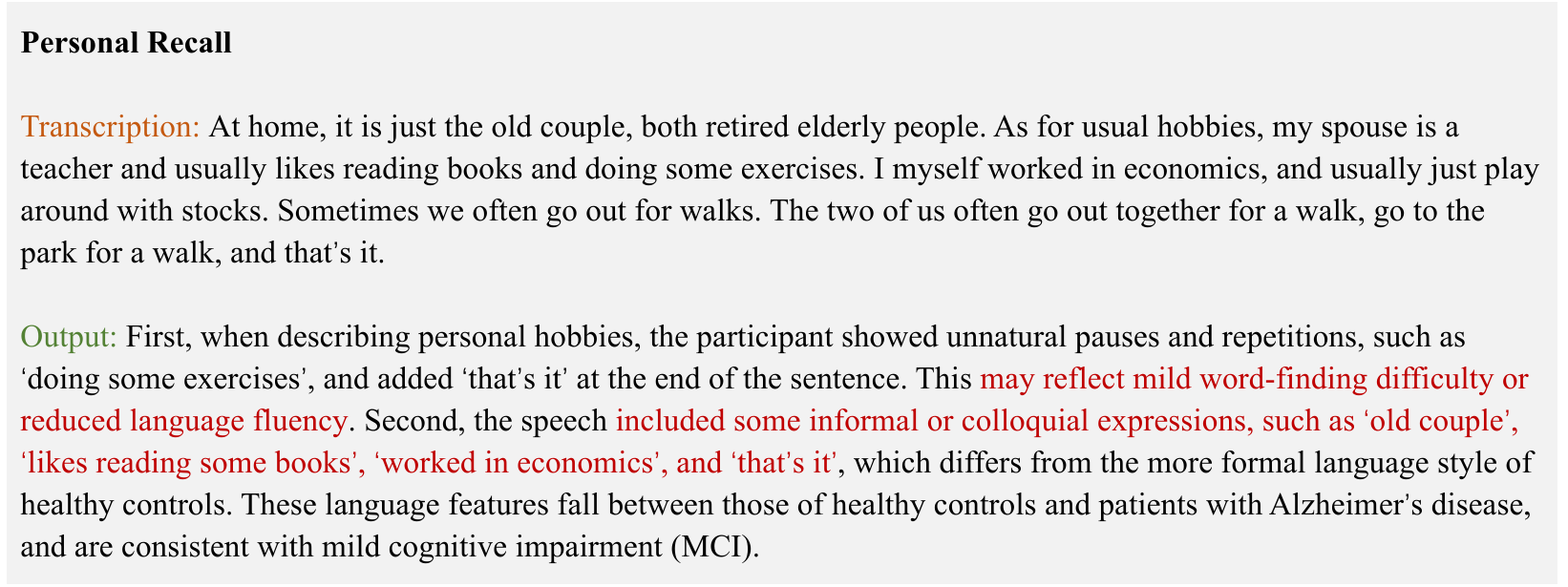}
        \caption{Failure case on the Personal Recall task. The cognitive status label is HC, whereas the model incorrectly predicts MCI. The red text highlights the model’s erroneous rationale: normal colloquial expressions, mild repetition, and natural sentence-final phrases are misinterpreted as evidence of word-finding difficulty or reduced fluency. This reflects an over-pathologisation of ordinary spontaneous speech.}
    \end{subfigure}
    \caption{Two representative failure cases of generated explanations on the CTD task and PR task. Each example includes the ASR transcription of the participant's speech, the system's generated explanation output, and the ground truth category. The original transcription and output are in Chinese, and have been translated to English here.}
    \label{fig:fail}
\end{figure}

Supplementary Figure~\ref{fig:fail} shows two representative failure cases.
Failures happen in cases where the model produces an incorrect classification. In the CTD example, the participant was in an AD case, but the model predicted HC. The explanation focused on the participant’s ability to describe several scene elements, but it treated inaccurate or confused descriptions as valid evidence of preserved language ability.
In the PR example, where the participant was HC but classified as MCI, the explanation over-interpreted informal expressions and sentence endings as signs of impairment, although these expressions can also appear in normal spontaneous speech. 
This failure case suggests that explanation quality and classification correctness are related, and that low-quality explanations may help clinicians identify predictions that require more careful review. Together, these examples indicate that the model can generate useful and clinically relevant explanations, but it may still produce unsupported or over-amplified evidence in some cases. Therefore, generated explanations should be used as auxiliary information rather than clinical conclusions.

\subsection*{Prompts}
\label{supp:prompt}

\begin{figure}[h]
    \centering
    \begin{subfigure}{\linewidth}
        \centering
        \includegraphics[width=1.0\linewidth]{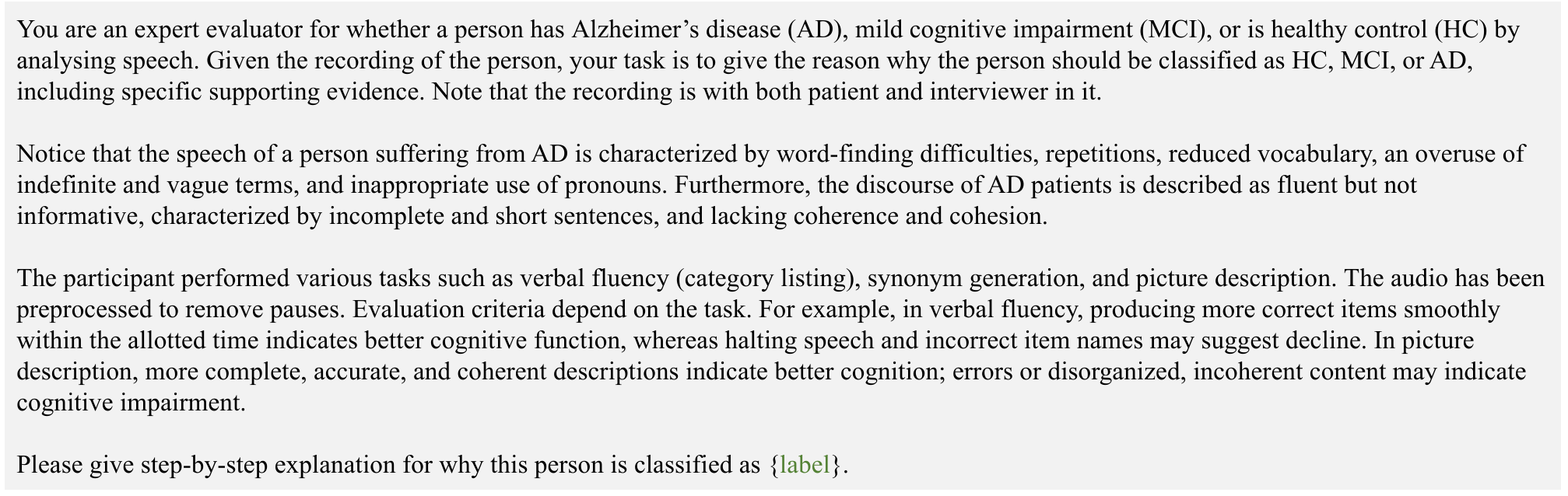}
        \caption{Prompt for Stage 1 on the NCMMSC-AD dataset.}
    \end{subfigure}
    \begin{subfigure}{\linewidth}
        \centering
        \includegraphics[width=1.0\linewidth]{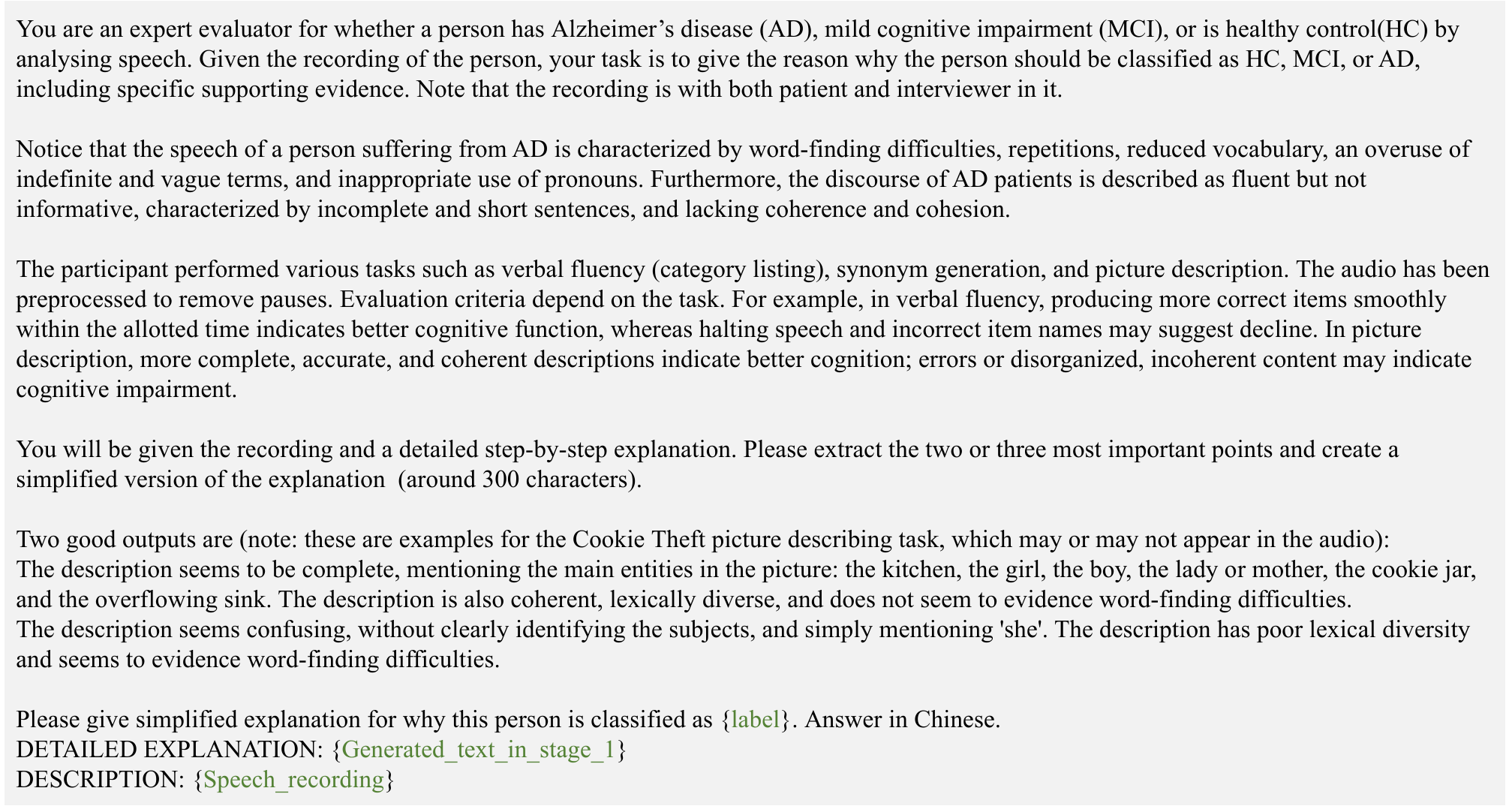}
        \caption{Prompt for Stage 2 on the NCMMSC-AD dataset.}
    \end{subfigure}
    \caption{Prompt templated for generating explanation texts using Gemini-2.5-Flash. The generation process employs a two-stage pipeline to ensure thorough analysis and concise output: Stage 1 guides detailed step-by-step analysis of speech characteristics, while Stage 2 extracts the most salient points from the detailed explanation. The prompt is original in Chinese and translated into English here.}
    \label{fig:prompt}
\end{figure}

Supplementary Figure~\ref{fig:prompt} presents the prompt template used to generate training target explanations through the two-stage pipeline with Gemini-2.5-Flash. The prompts are carefully structured to elicit clinically relevant reasoning that captures both acoustic and linguistic markers of cognitive impairment. Each prompt begins with a clear task definition positioning the model as an expert evaluator determining whether a person has AD, MCI, or is a healthy control based on speech analysis, or AD/HC for the ADReSS dataset. The prompts include comprehensive descriptions of speech characteristics commonly observed in AD patients, such as word-finding difficulties, repetitions, reduced vocabulary, overuse of indefinite and vague terms, inappropriate pronoun usage, and discourse features like fluent but non-informative speech with incomplete or short sentences lacking coherence. These descriptions serve as reference criteria that guide the model's attention toward clinically meaningful features rather than superficial speech patterns. Additionally, the prompts provide context about the specific cognitive tasks being performed along with task-specific evaluation criteria.

The two-stage generation process serves distinct but complementary purposes. Stage 1 prompts request detailed step-by-step explanations that systematically analyse the speech recording, encouraging thorough examination of multiple dimensions, including lexical diversity, semantic coherence, discourse organisation, and acoustic features. This comprehensive analysis covers a broad range of acoustic and linguistic features relevant to cognitive impairment. Stage 2 prompts the model to distil the detailed analysis into a simplified version, extracting only the two or three most important cognitive-relevant points, creating concise explanations suitable for clinical use. Stage 2 prompts additionally include example outputs demonstrating the desired explanation format and level of detail, providing the model with concrete templates to emulate. This two-stage approach balances thoroughness in feature identification with practical utility, producing explanations that are both comprehensive in their cognitive-relevant foundation and accessible in their final presentation, thereby ensuring that generated training targets effectively teach the SpeechLLM system to identify and articulate clinically meaningful speech patterns associated with cognitive impairment.

\begin{figure}
    \centering
    \includegraphics[width=1.0\linewidth]{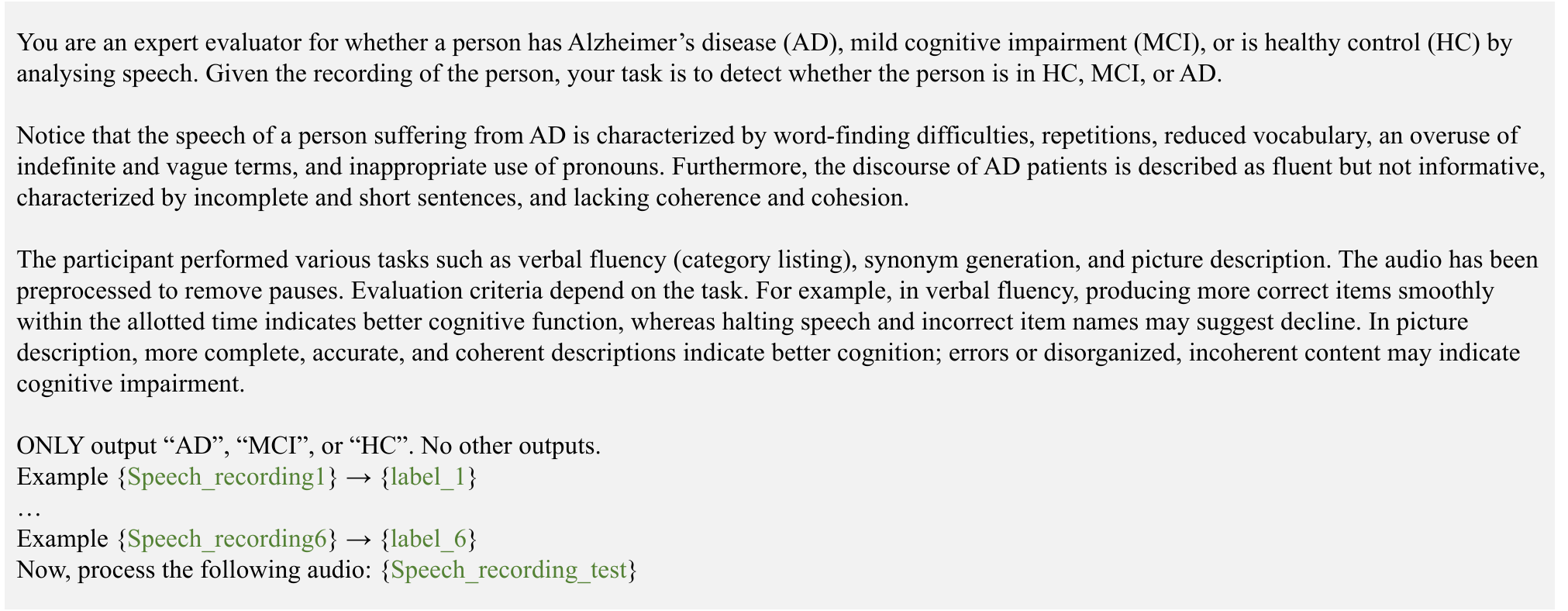}
    \caption{Prompt templated for Gemini-few-shot baseline, with NCMMSC-AD shown as an example. The prompt contains the same general clinical and task descriptions as the explanation generation prompt, but is adapted for direct cognitive status classification. Six in-context examples are provided, including two examples from each category (HC, MCI, and AD). The prompt is original in Chinese and translated into English here.}
    \label{fig:few-shot}
\end{figure}

Supplementary Figure~\ref{fig:few-shot} presents the prompt template used for the Gemini-few-shot baseline, with NCMMSC-AD shown as a representative example. The first part of the prompt follows the same general structure as the explanation generation prompt. Unlike the explanation generation prompt, this prompt was designed for direct cognitive status classification rather than rationale generation. For NCMMSC-AD, the model was given six in-context examples, with two examples from each cognitive status category: HC, MCI, and AD. Each example consisted of a speech recording and its corresponding cognitive label. After these examples, the model was asked to classify the test speech sample and output only one of the three cognitive status categories: HC, MCI, or AD. This setting was used to evaluate whether a general-purpose multimodal language model could perform speech-based cognitive classification without task-specific fine-tuning.

\begin{figure}
    \centering
    \includegraphics[width=1.0\linewidth]{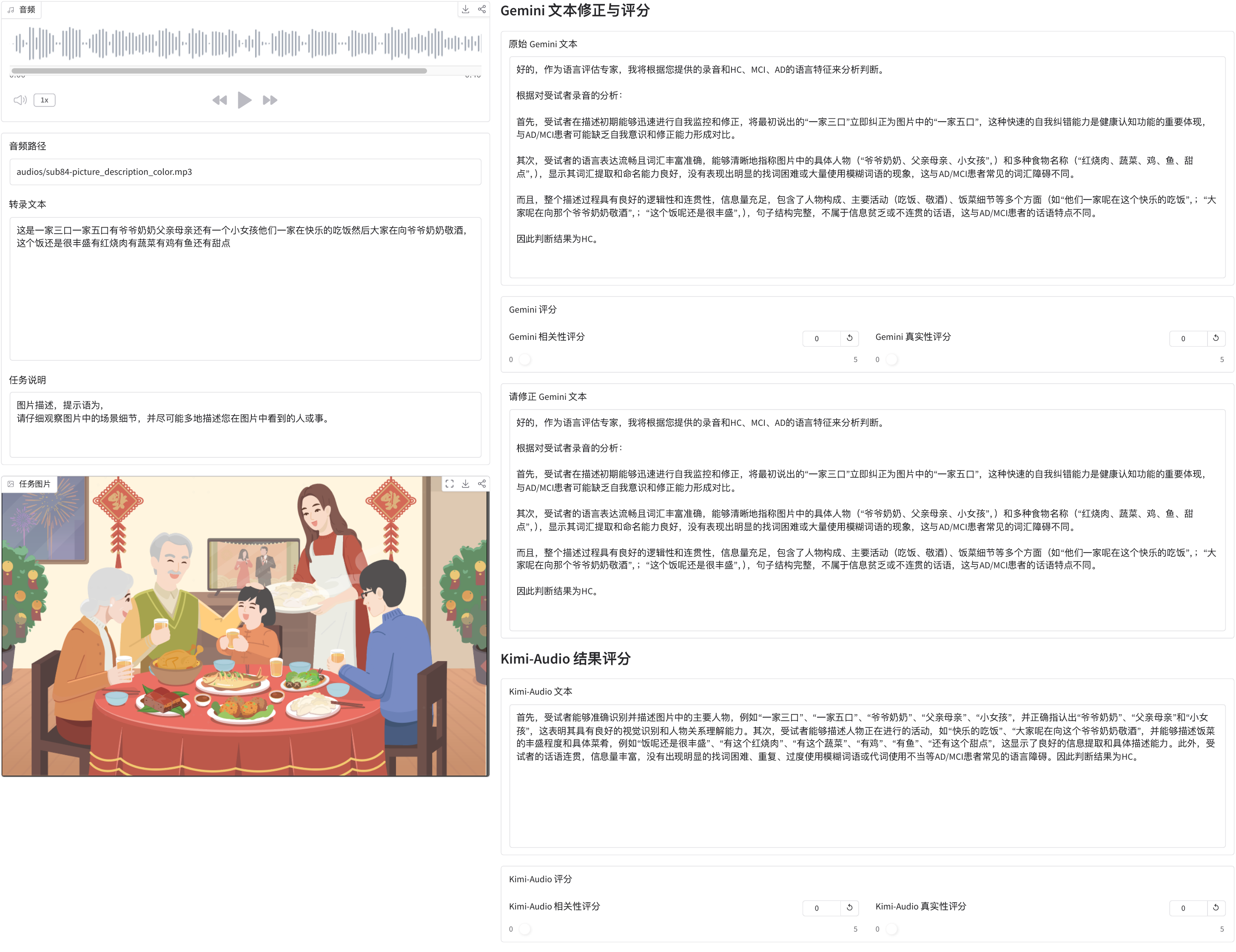}
    \caption{Web interface for clinician evaluation of generated explanations. The left panel presents the audio player, audio path, transcription, task instruction, and task image when applicable. The right panel presents the original Gemini-generated explanation, scoring fields for the Gemini explanation, an editable text box for clinician correction, the Kimi-Audio-generated explanation, and scoring fields for the Kimi-Audio explanation. The interface was implemented in Chinese and deployed on a local server. A default value of 0 indicates that a rating has not yet been assigned; valid clinician ratings range from 1 to 5.}
    \label{fig:webeval}
\end{figure}

\begin{sidewaystable}[t]
\centering

\caption{Classification performance (accuracy and AUROC) and explanation quality (BERT-Score for semantic similarity and hit rate (HR) for keyword matching) across datasets and speech tasks. SSL-Model baseline, Gemini-few-shot baseline, Text-only-LLM baseline, alongside our proposed SpeechLLM approach with respective SALMONN and Kimi-Audio backbones (Ours-SALMONN and Ours-Kimi-Audio) are listed. ADR represents the ADReSS corpus and NCM for NCMMSC-AD. Each model was trained and tested separately on individual datasets/tasks without cross-domain training. Results reported as mean $\pm$ standard deviation across the five different seeds. For accuracy and AUROC, the 95\% subject-level bootstrap confidence intervals (CIs) are shown in square brackets. The last column (Avg.) reports the average over six dataset/task conditions. The highest metrics are marked in bold.}

\label{table: overall}

\begin{tabular}{@{}ccccccccc@{}}
\toprule
                            &                                  & \multirow{2}{*}{\textbf{ADR}} & \multirow{2}{*}{\textbf{NCM}} & \multicolumn{4}{c}{\textbf{PUTH-AD}}                               & \multirow{2}{*}{\textbf{Avg.}}  \\
                            &                                  &                               &                               & \textbf{AFT}   & \textbf{FPD}   & \textbf{CTD}   & \textbf{PR}     &                                 \\ \midrule
\multirow{10}{*}{Accuracy}  & \multirow{2}{*}{SSL-Model}       & 0.831 ± 0.010                 & 0.847 ± 0.025                 & 0.459 ± 0.055  & 0.441 ± 0.031  & 0.459 ± 0.040  & 0.497 ± 0.051   & \multirow{2}{*}{0.589}          \\
                            &                                  & [0.741, 0.913]                & [0.787, 0.903]                & [0.352, 0.566] & [0.331, 0.552] & [0.355, 0.562] & [0.390, 0.603]  &                                 \\
                            & \multirow{2}{*}{Gemini-few-shot} & 0.637 ± 0.081                 & 0.392 ± 0.033                 & 0.269 ± 0.036  & 0.376 ± 0.045  & 0.376 ± 0.046  & 0.462 ± 0.077   & \multirow{2}{*}{0.418}          \\
                            &                                  & [0.561, 0.710]                & [0.326, 0.455]                & [0.176, 0.366] & [0.286, 0.466] & [0.279, 0.479] & [0.376, 0.552]  &                                 \\
                            & \multirow{2}{*}{Text-only-LLM}   & 0.802 ± 0.029                 & 0.763 ± 0.018                 & 0.453 ± 0.034  & 0.461 ± 0.049  & 0.534 ± 0.041  & 0.534 ± 0.047   & \multirow{2}{*}{0.591}          \\
                            &                                  & [0.724, 0.876]                & [0.694, 0.829]                & [0.354, 0.554] & [0.358, 0.565] & [0.435, 0.638] & [0.431, 0.635]] &                                 \\
                            & \multirow{2}{*}{Ours-SALMONN}    & 0.882 ± 0.011                 & 0.825 ± 0.016                 & 0.462 ± 0.037  & 0.466 ± 0.052  & 0.528 ± 0.037  & 0.452 ± 0.020   & \multirow{2}{*}{0.602}          \\
                            &                                  & [0.817, 0.938]                & [0.766, 0.881]                & [0.362, 0.562] & [0.369, 0.562] & [0.431, 0.621] & [0.359, 0.545]  &                                 \\
                            & \multirow{2}{*}{Ours-Kimi-Audio} & 0.893 ± 0.016                 & 0.857 ± 0.023                 & 0.553 ± 0.016  & 0.507 ± 0.029  & 0.553 ± 0.041  & 0.638 ± 0.025   & \multirow{2}{*}{\textbf{0.666}} \\
                            &                                  & [0.825, 0.952]                & [0.807, 0.904]                & [0.427, 0.677] & [0.404, 0.612] & [0.442, 0.662] & [0.523, 0.750]  &                                 \\ \midrule
\multirow{8}{*}{AUROC}      & \multirow{2}{*}{SSL-Model}       & 0.876 ± 0.009                 & 0.958 ± 0.008                 & 0.589 ± 0.029  & 0.615 ± 0.037  & 0.633 ± 0.014  & 0.652 ± 0.040   & \multirow{2}{*}{0.721}          \\
                            &                                  & [0.786, 0.951]                & [0.929, 0.981]                & [0.492, 0.689] & [0.531, 0.699] & [0.535, 0.728] & [0.544, 0.753]  &                                 \\
                            & \multirow{2}{*}{Text-only-LLM}   & 0.911 ± 0.025                 & 0.872 ± 0.010                 & 0.594 ± 0.026  & 0.588 ± 0.050  & 0.709 ± 0.025  & 0.695 ± 0.049   & \multirow{2}{*}{0.728}          \\
                            &                                  & [0.851, 0.960]                & [0.816, 0.921]                & [0.512, 0.687] & [0.533, 0.723] & [0.628, 0.793] & [0.605, 0.790]  &                                 \\
                            & \multirow{2}{*}{Ours-SALMONN}    & 0.922 ± 0.013                 & 0.940 ± 0.008                 & 0.603 ± 0.024  & 0.651 ± 0.028  & 0.722 ± 0.027  & 0.618 ± 0.046   & \multirow{2}{*}{0.742}          \\
                            &                                  & [0.859, 0.971]                & [0.905, 0.968]                & [0.514, 0.691] & [0.559, 0.736] & [0.637, 0.801] & [0.522, 0.709]  &                                 \\
                            & \multirow{2}{*}{Ours-Kimi-Audio} & 0.921 ± 0.013                 & 0.943 ± 0.004                 & 0.675 ± 0.031  & 0.673 ± 0.011  & 0.681 ± 0.032  & 0.770 ± 0.022   & \multirow{2}{*}{\textbf{0.777}} \\
                            &                                  & [0.849, 0.980]                & [0.906, 0.973]                & [0.587, 0.763] & [0.584, 0.763] & [0.572, 0.792] & [0.674, 0.875]  &                                 \\ \midrule
\multirow{2}{*}{BERT-Score} & Ours-SALMONN                     & 0.754 ± 0.004                 & 0.544 ± 0.010                 & 0.546 ± 0.003  & 0.551 ± 0.008  & 0.561 ± 0.007  & 0.556 ± 0.011   & 0.585                           \\
                            & Ours-Kimi-Audio                  & 0.772 ± 0.001                 & 0.727 ± 0.007                 & 0.731 ± 0.012  & 0.743 ± 0.009  & 0.752 ± 0.004  & 0.742 ± 0.004   & \textbf{0.744}                  \\ \midrule
\multirow{2}{*}{HR}         & Ours-SALMONN                     & 0.567 ± 0.048                 & 0.055 ± 0.029                 & 0.000 ± 0.000  & 0.006 ± 0.004  & 0.012 ± 0.007  & 0.012 ± 0.010   & 0.108                           \\
                            & Ours-Kimi-Audio                  & 0.697 ± 0.037                 & 0.580 ± 0.032                 & 0.598 ± 0.053  & 0.570 ± 0.043  & 0.569 ± 0.022  & 0.613 ± 0.026   & \textbf{0.604}                  \\ \bottomrule
\end{tabular}
\end{sidewaystable}

\begin{sidewaystable}[t]
\centering
\caption{Ablation study comparing classification-only, generation-only, and multi-task training (Ours) approaches across SALMONN and Kimi-Audio backbones. Results reported as mean $\pm$ standard deviation across the five different seeds. The Avg. column reports the average over six dataset/task conditions. Highest average metrics marked in bold.}
\label{table: ablation}
% \resizebox{\linewidth}{!}{%
\begin{tabular*}{\linewidth}{@{}ccccccccc@{}}
\toprule
\multicolumn{2}{c}{\multirow{2}{*}{\textbf{SALMONN}}} & \multirow{2}{*}{\textbf{ADR}} & \multirow{2}{*}{\textbf{NCM}} & \multicolumn{4}{c}{\textbf{PUTH-AD}}                          & \multirow{2}{*}{\textbf{Avg.}} \\
\multicolumn{2}{c}{}                                  &                               &                               & \textbf{AFT}  & \textbf{FPD}  & \textbf{CTD}  & \textbf{PR}   &                                \\ \midrule
\multirow{3}{*}{Accuracy}                & Classification              & 0.815 ± 0.025                 & 0.809 ± 0.039                 & 0.469 ± 0.025 & 0.469 ± 0.020 & 0.566 ± 0.013 & 0.476 ± 0.023 & 0.600                          \\
                        & Generation                  & 0.755 ± 0.063                 & 0.442 ± 0.040                 & 0.379 ± 0.048 & 0.438 ± 0.033 & 0.450 ± 0.038 & 0.359 ± 0.047 & 0.471                          \\
                        & Ours-SALMONN                & 0.882 ± 0.011                 & 0.825 ± 0.016                 & 0.462 ± 0.037 & 0.466 ± 0.052 & 0.528 ± 0.037 & 0.452 ± 0.020 & \textbf{0.602}                 \\ \midrule
\multirow{2}{*}{AUROC}                   & Classification              & 0.885 ± 0.027                 & 0.943 ± 0.008                 & 0.638 ± 0.032 & 0.656 ± 0.010 & 0.720 ± 0.014 & 0.601 ± 0.072 & 0.740                          \\
                        & Ours-SALMONN                & 0.922 ± 0.013                 & 0.940 ± 0.008                 & 0.603 ± 0.024 & 0.651 ± 0.028 & 0.722 ± 0.027 & 0.618 ± 0.046 & \textbf{0.742}                 \\ \midrule
\multirow{2}{*}{BERT-Score}              & Generation                  & 0.756 ± 0.006                 & 0.716 ± 0.004                 & 0.702 ± 0.006 & 0.716 ± 0.006 & 0.722 ± 0.006 & 0.709 ± 0.006 & \textbf{0.720}                 \\
                        & Ours-SALMONN                & 0.754 ± 0.004                 & 0.544 ± 0.010                 & 0.546 ± 0.003 & 0.551 ± 0.008 & 0.561 ± 0.007 & 0.556 ± 0.011 & 0.585                          \\ \midrule
\multirow{2}{*}{HR}                      & Generation                  & 0.576 ± 0.045                 & 0.396 ± 0.049                 & 0.403 ± 0.041 & 0.380 ± 0.015 & 0.335 ± 0.033 & 0.447 ± 0.027 & \textbf{0.423}                 \\
                        & Ours-SALMONN                & 0.567 ± 0.048                 & 0.055 ± 0.029                 & 0.000 ± 0.000 & 0.006 ± 0.004 & 0.012 ± 0.007 & 0.012 ± 0.010 & 0.108                          \\
\midrule \midrule
\multicolumn{2}{c}{\multirow{2}{*}{\textbf{Kimi-Audio}}} & \multirow{2}{*}{\textbf{ADR}} & \multirow{2}{*}{\textbf{NCM}} & \multicolumn{4}{c}{\textbf{PUTH-AD}}                          & \multirow{2}{*}{\textbf{Avg.}} \\
\multicolumn{2}{c}{}                                     &                               &                               & \textbf{AFT}  & \textbf{FPD}  & \textbf{CTD}  & \textbf{PR}   &                                \\ \midrule
\multirow{3}{*}{Accuracy}         & Classification       & 0.754 ± 0.023                 & 0.810 ± 0.014                 & 0.530 ± 0.017 & 0.507 ± 0.025 & 0.507 ± 0.025 & 0.626 ± 0.039 & 0.622                          \\
                                  & Generation           & 0.480 ± 0.021                 & 0.501 ± 0.023                 & 0.442 ± 0.059 & 0.465 ± 0.017 & 0.576 ± 0.045 & 0.538 ± 0.029 & 0.500                          \\
                                  & Ours-Kimi-Audio      & 0.893 ± 0.016                 & 0.857 ± 0.023                 & 0.553 ± 0.016 & 0.507 ± 0.029 & 0.553 ± 0.041 & 0.638 ± 0.025 & \textbf{0.666}                 \\ \midrule
\multirow{2}{*}{AUROC}            & Classification       & 0.829 ± 0.019                 & 0.923 ± 0.014                 & 0.649 ± 0.007 & 0.702 ± 0.010 & 0.644 ± 0.017 & 0.747 ± 0.064 & 0.749                          \\
                                  & Ours-Kimi-Audio      & 0.921 ± 0.013                 & 0.943 ± 0.004                 & 0.675 ± 0.031 & 0.673 ± 0.011 & 0.681 ± 0.032 & 0.770 ± 0.022 & \textbf{0.777}                 \\ \midrule
\multirow{2}{*}{BERT-Score}       & Generation           & 0.778 ± 0.002                 & 0.739 ± 0.005                 & 0.734 ± 0.011 & 0.753 ± 0.006 & 0.753 ± 0.002 & 0.746 ± 0.002 & \textbf{0.750}                 \\
                                  & Ours-Kimi-Audio      & 0.772 ± 0.001                 & 0.727 ± 0.007                 & 0.731 ± 0.012 & 0.743 ± 0.009 & 0.752 ± 0.004 & 0.742 ± 0.004 & 0.744                          \\ \midrule
\multirow{2}{*}{HR}               & Generation           & 0.666 ± 0.051                 & 0.619 ± 0.035                 & 0.586 ± 0.060 & 0.576 ± 0.012 & 0.561 ± 0.040 & 0.652 ± 0.025 & \textbf{0.610}                 \\
                                  & Ours-Kimi-Audio      & 0.697 ± 0.037                 & 0.580 ± 0.032                 & 0.598 ± 0.053 & 0.570 ± 0.043 & 0.569 ± 0.022 & 0.613 ± 0.026 & 0.604                          \\ \bottomrule
\end{tabular*}
% }
\end{sidewaystable}

\begin{table}[t]
\caption{Comparison of separate training versus cross-domain joint training with Kimi-Audio backbone. Joint training data includes ADReSS, NCMMSC-AD, and PUTH-AD Personal Recall data. “Unseen” denotes PUTH-AD task subsets excluded from joint training, which are evaluated on the held-out PUTH-AD test subjects for generalisation assessment. Voting represents the result of hard voting across all PUTH-AD tasks. Results reported as mean $\pm$ standard deviation across the five different seeds.}
\label{tab: generalisation}
\centering
\begin{tabular}{@{}ccccccc@{}}
\toprule
\multirow{2}{*}{Dataset} & \multirow{2}{*}{Speech Task} &              & \multicolumn{4}{c}{Separate Training}                         \\ \cmidrule(l){3-7} 
                         &                              & Unseen       & Acc           & AUROC         & BERT          & HR            \\ \midrule
ADReSS                   & -                            & -            & 0.893 ± 0.016 & 0.921 ± 0.013 & 0.772 ± 0.001 & 0.697 ± 0.037 \\
NCMMSC-AD                & -                            & -            & 0.857 ± 0.023 & 0.943 ± 0.004 & 0.727 ± 0.007 & 0.580 ± 0.032 \\
\multirow{5}{*}{PUTH-AD} & AFT                          & -            & 0.553 ± 0.016 & 0.675 ± 0.031 & 0.731 ± 0.012 & 0.598 ± 0.053 \\
                         & FPD                          & -            & 0.507 ± 0.029 & 0.673 ± 0.011 & 0.743 ± 0.009 & 0.570 ± 0.043 \\
                         & CTD                          & -            & 0.553 ± 0.041 & 0.681 ± 0.032 & 0.752 ± 0.004 & 0.569 ± 0.022 \\
                         & PR                           & -            & 0.638 ± 0.025 & 0.770 ± 0.022 & 0.742 ± 0.004 & 0.613 ± 0.026 \\ \cmidrule(l){2-7} 
                         & Voting                       &              & 0.634 ± 0.050 & -             & -             & -             \\ \midrule \midrule
\multirow{2}{*}{Dataset} & \multirow{2}{*}{Speech Task} & \multicolumn{5}{c}{Joint training}                                           \\ \cmidrule(l){3-7} 
                         &                              & Unseen       & Acc           & AUROC         & BERT          & HR            \\ \midrule
ADReSS                   & -                            &              & 0.862 ± 0.027 & 0.907 ± 0.026 & 0.773 ± 0.004 & 0.701 ± 0.023 \\
NCMMSC-AD                & -                            &              & 0.837 ± 0.033 & 0.941 ± 0.008 & 0.740 ± 0.002 & 0.616 ± 0.029 \\
\multirow{5}{*}{PUTH-AD} & AFT                          & $\checkmark$ & 0.538 ± 0.036 & 0.651 ± 0.019 & 0.736 ± 0.003 & 0.572 ± 0.063 \\
                         & FPD                          & $\checkmark$ & 0.550 ± 0.021 & 0.706 ± 0.011 & 0.749 ± 0.005 & 0.513 ± 0.048 \\
                         & CTD                          & $\checkmark$ & 0.576 ± 0.040 & 0.730 ± 0.046 & 0.748 ± 0.002 & 0.604 ± 0.017 \\
                         & PR                           &              & 0.630 ± 0.039 & 0.750 ± 0.044 & 0.744 ± 0.004 & 0.613 ± 0.052 \\ \cmidrule(l){2-7} 
                         & Voting                       &              & 0.700 ± 0.029 & -             & -             & -             \\ \bottomrule
\end{tabular}
% }
\end{table}

\end{document}